\documentclass[useAMS,usenatbib]{mn2e}
\usepackage{graphicx,amssymb,amsmath,times}
\title{On the point source approximation of nearby cosmic-ray sources}
\author[Satyendra Thoudam]{Satyendra Thoudam\thanks{E-mail: s.thoudam@astro.ru.nl} and J\"{o}rg R. H\"{o}randel\\
Department of Astrophysics, IMAPP, Radboud University Nijmegen, P.O. Box 9010, 6500 GL Nijmegen, The Netherlands}

\begin{document}
\date{}
\pagerange{}
\maketitle
\label{firstpage}
\begin{abstract}
In this paper, we check in detail the validity of the widely adopted point source approximation for nearby cosmic-ray (CR) sources. Under an energy independent escape model for CRs from the sources, we show that for young nearby sources, the point source approximation breaks down at lower energies and the CR spectrum depends on the size and the morphology of the source. When applied to the nearby supernova remnants (SNRs), we find that the approximation breaks down for some of the individual remnants like the Vela, but interestingly it still holds good for their combined total spectrum at the Earth. Moreover, we also find that the results obtained under this simple approximation are quite different from those calculated under an energy dependent escape model which is favored by diffusive shock acceleration models inside SNRs. Our study suggests that if SNRs are the main sources of CRs in our Galaxy, then the commonly adopted point source model (with an energy independent escape scenario) appears flawed for CR studies from the nearby SNRs.
\end{abstract}

\section{Introduction}
Cosmic rays (CRs) with energies below the knee region ($\sim 3\times 10^{15}$ eV) are considered to be of galactic origin. Although the exact nature of their sources are not known, the most favorable candidates are the supernova remnants (SNRs). They   are known to occur in our Galaxy at the rate of $\sim 1/30$ yr$^{-1}$ with each explosion releasing a total kinetic energy of $\sim 10^{51}$ ergs. If approximately $10\%$ of this energy is converted into CRs, then the total power release is sufficient to maintain the CR energy density  in our Galaxy which is measured to be around $1$ eV cm$^{-3}$.

It is also now theoretically established that SNR shock waves can accelerate CRs up to energies close to the knee by diffusive shock acceleration (DSA) mechanism (Bell 1978, Blandford $\&$ Eichler 1987). In a simple planar shocks model, such a mechanism naturally leads to a power-law spectrum of the form $E^{-\Gamma}$ with the exponent $\Gamma=2$ for strong shocks. This value  is found to be in good agreement with the radio observations of SNRs (Green 2009). In addition, direct evidence for the presence of high energy particles up to few TeVs ($1$ TeV= $10^{12}$ eV) inside SNRs comes from the detections of non-thermal X-rays (see e.g., Parizot et al. 2006, Bamba et al. 2006) and high energy TeV $\gamma$-rays from several SNRs (e.g., Aharonian et al. 2006, 2008a, 2008b, 2008d, Albert et al. 2007). The non-thermal X-rays are best explained as synchrotron emission from high energy electrons while the origin of the TeV $\gamma$-rays is still not certain between the leptonic (via inverse compton process) and the hadronic scenarios (via $\pi^0$ decays). If the high energy $\gamma$-rays are of hadronic origin as indicated by the recent observations of several SNRs by the FERMI experiment (Abdo et al. 2009, 2010a, b, c), then the measured $\gamma$-rays can provide direct informations about the spectral shape of the primary particles. But, TeV measurements made by the new generation Cherenkov telescopes like the HESS, MAGIC and VERITAS have found that many SNRs show $\Gamma\sim (2.3-2.7)$ which is steeper than the expectations from DSA theory. The discrepancy becomes even more severe if we compare with the results of non-linear DSA theory which predicts a spectrum flatter than $\Gamma=2$ (Caprioli et al. 2010 and references therein). Although this discrepancy is still not yet fully understood, for our study we will assume that SNRs are the main sources of CRs in our Galaxy.

Quite often, theoretical studies on the propagation of CRs assume the sources to be stationary and continuously distributed in the Galaxy. This simple assumption seems reasonable for calculating the Galactic average CR properties and for understanding the  diffuse radiations produced by the interaction of CRs in the interstellar medium (ISM). But, for CR studies in the vicinity of the sources where the influence of the source is expected to dominate over the background produced by the distant sources, the discrete nature of the sources (both in space and time) may become important. For instance, in the study of gamma-ray emissions from the environment of the sources or from molecular clouds associated with them, the emission can be strongly dependent on the age and the distance of the source as discussed in Aharonian $\&$ Atoyan 1996, Gabici et al. 2009, Casanova et al. 2010.

Similarly, for the study of CRs observed at the Earth, the uniform source distribution looks proper only for the distant sources but for the nearby sources, a more reasonable treatment would be to consider the discrete nature of the sources. For CR electrons at few TeV energies, such treatment seems even more important because of their fast energy loss rate. Electrons with energies greater than $\sim 1$ TeV cannot travel distances more than $\sim 1$ kpc in the Galaxy through diffusive propagation before they lost all their energies. Therefore, high energy electrons from distant and old sources may not reach the Earth effectively and it is possible that most of the TeV electrons that we observe are mostly produced by few young nearby sources (Shen 1970, Atoyan et al. 1995, Kobayashi et al. 2004, Delahaye et al. 2010 etc.).  Also for the CR protons and other nuclear species which do not suffer significant losses (the typical nuclear fragmentation loss time scale in our Galaxy $\sim 10^7$ yr) and for which we expect a strong background from the distant sources, the discrete treatment of the nearby sources can still be important especially at higher energies (Strong $\&$ Moskalenko 2001, B\"ushing et al. 2005, Erlykin $\&$ Wolfendale 2006 and references therein). This is because high energy CRs diffuse relatively faster compared to the lower energy ones and hence, they are expected to produce stronger fluctuations on their observable properties like the spectrum and the anisotropy (Thoudam 2008). Moreover, at these energies, the contribution from  the recent sources may dominate and the effect of discreteness in time may also become important (Taillet et al. 2004).

In most of the studies mentioned above, the discrete sources are assumed to be point-like, thereby neglecting their finite size and the morphology. At first sight, the point source approximation seems reasonable for sources whose size $s\ll r$, the distance from the Earth. But, for those whose size is comparable to the distance, the point source approximation may break down and it looks more appropriate to take their size and morphology into account. Under the standard DSA theory, CRs are confined within the SNRs due to the strong magnetic turbulence generated by the CRs themselves and therefore, it is reasonable to assume that CRs remain confined as long as the shocks remain strong enough to act as an efficient accelerator. For a typical interstellar matter density of $1$ H cm$^{-3}$, the confinement last until the SNR age around $10^5$ yr (Berezhko $\&$ V\"olk 2000). In reality, an energy dependent confinement/escape scenario is expected (see e.g. Ptuskin $\&$ Zirakashvili 2005, Caprioli et al. 2009). Using the Sedov relation between the SNR age and the shock radius, if we assume an initial shock velocity of $10^9$ cm s$^{-1}$, we can roughly estimate that at the age of $10^5$ yr the remnant expands to a size of radius around $100$ pc. Such a size is comparable to the distance of some of the nearest SNRs like the Geminga and the Loop1. The distance to the Geminga is estimated to be $\sim 157$ pc (Caraveo et al. 1996) and that to the Loop1 as $\sim 170$ pc (Egger $\&$ Aschenbach 1995).

The argument just mentioned is purely based on the geometrical consideration, i.e., the source size compared to its distance and we have not considered any possible effects due to the propagation of CRs in the Galaxy. It is now well accepted that CRs undergo diffusive motion due to scattering by the magnetic field irregularities and the self excited hydromagnetic waves in the ISM. Measurements of secondary-to-primary (s/p) ratios like the boron-to-carbon indicate that the diffusion is energy dependent with the diffusion coefficient increasing with energy (see e.g. the experiments listed in Stephens $\&$ Streitmatter 1998). If we also take into account such an energy dependent diffusion, the validity of the point source approximation may become somewhat relaxed for the high energy particles, i.e, it may still represent a good approximation even for the nearby sources at higher energies. We will discuss this in detail later in the paper.

Recently, Ohira et al. 2011 highlighted the importance of the finite source size in the study of gamma-ray emission from SNRs interacting with molecular clouds. They claimed that the observations of different gamma-ray spectra from four SNRs (W51C, W28, W44 and IC 443) by the FERMI experiment could be an effect of finite size of the SNRs. In this paper, we will investigate the importance of the source size for the nearby SNRs considering the CR spectrum expected at the Earth. Although SNRs can have complex morphologies, that too different from each other, for simplicity we will consider a spherical geometry for our study. In one part, we will consider an energy \textit{independent} escape of CRs from the SNRs. This is discussed in sections 2 and 3. In another part of our study, we will investigate the energy \textit{dependent} escape model under which CRs of different energies are assumed to escape at different times. This study is given in section 4.  Then in section 5, we apply our study to the nearby known SNRs and compare the results obtained under the different source models. Finally in section 6, we present an overview of our results and discuss their implications.

\section {CR proton spectrum from an SNR}
In the diffusion model, neglecting losses due to interactions in the ISM, the propagation of CR protons in the Galaxy can be described by (see e.g. Gaisser 1990 and references therein),
\begin{equation}
\nabla\cdot(D\nabla N_p)+Q_p=\frac{\partial N_p}{\partial t}
\end{equation} 
where $N_p(\textbf{r},E,t)$ is the differential proton density, $E$ is the kinetic energy, $D(E)$ is the diffusion coefficient and $Q_p(\textbf{r},E,t)$ is the source term, i.e. the proton production rate from the SNR. In Eq. (1), we also neglect other effects which are relevant mostly below a few GeVs like the convection due to the Galactic wind and the reacceleration by the interstellar turbulence. We assume the diffusion coefficient to be spatially constant throughout the Galaxy and take $D(E)=D_0(E/E_0)^\delta$ for $E>E_0$, where $D_0=2.9\times 10^{28}$ cm$^2$ s$^{-1}$, $E_0=3$ GeV and $\delta=0.6$ (Thoudam 2008). These values are different from those given by models based on diffusive reacceleration. For instance, Trotta et al. 2011 give a value of $D_0\sim 8.3\times 10^{28}$  cm$^2$ s$^{-1}$ and $\delta\sim 0.3$. 

For sources within a distance of $\sim 1$ kpc from the Earth which are also our main interest here, Thoudam 2007 showed that the CR spectrum is not much affected by the presence of the Galactic boundaries. In fact, $D_0$ does depend on the boundaries and is proportional to the size of our Galactic halo (see e.g. Trotta et al. 2011). For our present study, we neglect such dependencies and solve Eq. (1) without imposing any boundary conditions. We then obtain the well known Green function $G_p(\textbf{r},\textbf{r}^\prime,t,t^\prime)$, i.e. the solution for a $\delta$-function source term $Q_p(\textbf{r},E,t)=\delta(\textbf{r}-\textbf{r}^\prime)\delta(t-t^\prime)$ as given below,
\begin{equation}
G_p(\textbf{r},\textbf{r}^\prime,t,t^\prime)=\frac{1}{8\left[\pi D(t-t^\prime)\right]^{3/2}}\mathrm{exp}\left[\frac{-(\textbf{r}^\prime -\textbf{r})^2}{4D(t-t^\prime)}\right]
\end{equation} 
The general solution of Eq. (1) can be then obtained using,
\begin{equation}
N_p(\textbf{r},E,t)=\int^{\infty}_{-\infty}d\textbf{r}^\prime\int^t_{-\infty}dt^\prime G_p(\textbf{r},\textbf{r}^\prime,t,t^\prime)Q_p(\textbf{r}^\prime,E,t^\prime)
\end{equation}
The source term in Eq. (3) can be written as,
\begin{equation}
Q_p(\textbf{r}^\prime,E,t^\prime)=q(\textbf{r}^\prime)q(E)q(t^\prime)
\end{equation}
where $q(E)$ is the source spectrum, i.e., $q(E)dE$ is the number of protons with energy between $E$ and $E+dE$ produced by the SNR. For this part of our study, we assume an energy independent escape of CRs from the SNR. We will first consider the burst-like injection of particles followed later by the continuous injection case. Later on, in section 4 we will discuss the energy dependent escape model.

If we assume that the burst-like emission of particles happen at time $t_0$, we can write the temporal source term as $q(t^\prime)=\delta(t^\prime-t_0)$. Then, using Eqs. (2) $\&$ (4) in Eq. (3), we get,
\begin{equation}
N_p(\textbf{r},E,t)=\frac{q(E)}{8\left[\pi D(t-t_0)\right]^{3/2}}\int^{\infty}_{-\infty}d\textbf{r}^\prime\mathrm{exp}\left[\frac{-(\textbf{r}^\prime -\textbf{r})^2}{4D(t-t_0)}\right]q(\textbf{r}^\prime)
\end{equation}
The proton intensity can be then calculated using the relation $I_p(E,t)\approx (c/4\pi)N_p(E,t)$, where $c$ is the velocity of light. To make our calculations simpler, hereafter we take $\textbf{r}=0$, i.e., we set the origin of the coordinate system at the position of the Earth. 

\subsection{Point source approximation}
If we assume the SNR to be a point source located at a distance $\textbf{r}_s$ from the Earth, we can write 
\begin{equation}
q(\textbf{r}^\prime)=\delta(\textbf{r}^\prime-\textbf{r}_s)
\end{equation}
Then, the proton density is obtained from Eq. (5) as, 
\begin{equation}
N_p(E,t)=\frac{q_p(E)}{8\left[\pi D(t-t_0)\right]^{3/2}}\mathrm{exp}\left[\frac{-r_s^2}{4D(t-t_0)}\right]
\end{equation}
Eq. (7) represents the most commonly adopted solution for CR spectrum from a nearby single source. For high energy particles for which the diffusion radius defined as $r_{diff}=\sqrt{4D(t-t_0)}$ is much larger than the distance to the SNR $r_s$, the exponential term in Eq. (7) tends to $1$ which implies,
\begin{equation}
N_p(E,t)\rightarrow\frac{q_p(E)}{8\left[\pi D(t-t_0)\right]^{3/2}}
\end{equation}
For a power-law source spectrum given by $q_p(E)=k_pE^{-\Gamma}$ and for $D(E)\propto E^\delta$, Eq. (8) shows that the spectrum of high energy protons reaching us follows $N_p(E)\propto E^{-\left(\Gamma+\frac{3}{2}\delta\right)}$. Particles with $r_{diff}>r_s$ are those which have already passed the Earth. Those with $r_{diff}<r_s$ are the ones which have not yet reached effectively due to their slower diffusion and therefore, their intensity is comparatively much suppressed.

\subsection{Spherical solid source}
Most of the SNRs are observed to roughly follow a spherical geometry and they come under three main categories: shell-type, plerion-type and composite-type. Shell-type SNRs show bright shell structure which expands into the ISM with velocities of $\sim (3-10)\times 10^8$ cm s$^{-1}$  (e.g. Cassiopeia A). Plerions also known as pulsar wind nebulae have filled center normally a pulsar powering high energy particles into the ISM (e.g. Crab Nebula). They do not show any shell-like features. Composite SNRs have both shell structure and filled center (e.g. IC443). The surface brightness of shell-type SNRs in radio as well as in X-rays are observed to peak near the surface whereas in plerions, it tends to increase towards the center. We can  expect that the high energy particles responsible for the radio and the X-ray emissions also follow a similar distribution within the remnant.

Let us now consider a spherical solid source. We believe that this source model roughly represents the plerions and the composite type SNRs. For this model, if $\textbf{r}_s$ denotes the position of the center of the SNR from the Earth and $\textbf{r}_0$ represents the position of the source CRs with respect to the SNR center, we can write $\textbf{r}^\prime$ in Eq. (5) as $\textbf{r}^\prime=\textbf{r}_s+\textbf{r}_0$ and then rewrite Eq. (5) as,
\begin{equation}
N_p(E,t)=\frac{q(E)}{8\left[\pi D(t-t_0)\right]^{3/2}}\int d\textbf{r}_0\,\mathrm{exp}\left[\frac{-({\textbf{r}_s+\textbf{r}_0})^2}{4D(t-t_0)}\right]q(\textbf{r}_0)
\end{equation}
In Eq. (9), the integral over the volume element in spherical geometry is given by,
\begin{equation*}
\int d\textbf{r}_0=\int^R_0 r_0^2\,dr_0\int^{\pi}_0\mathrm{sin\theta_0}\,d\theta_0\int^{2\pi}_0d\phi_0
\end{equation*}
where $R$ denotes the radius of the SNR. We take the source spectrum in this case as $q(E)=q_p(E)/V$ where $q_p(E)$ is the source spectrum we took in the case of the point source approximation (section 2.1) and $V=\frac{4}{3}\pi R^3$ represents the total SNR volume. We assume that CRs are uniformly distributed throughout the SNR volume before releasing into the ISM and we take the spatial source term as,
\begin{equation}
q(r_0)=\begin{cases}1, &\text{for $r_0\leq R$}\\
0, &\text{otherwise}\end{cases}
\end{equation}
Integrating Eq. (9) over $\theta_0$ and $\phi_0$, we get
\begin{eqnarray}
N_p(E,t)=\frac{q_p(E)}{r_sV\sqrt{\pi D(t-t_0)}}\mathrm{exp}\left[\frac{-r_s^2}{4D(t-t_0)}\right]\quad\quad\qquad\nonumber\\
\qquad\times\;\int^{R}_0r_0\,\mathrm{exp}\left[\frac{-r_0^2}{4D(t-t_0)}\right]\mathrm{sinh}\left(\frac{r_sr_0}{2D(t-t_0)}\right)dr_0
\end{eqnarray}
Using the properties $\mathrm{sinh}(x)\approx x$ and $e^x\rightarrow 1$ for very small $x$, it is easy to check that for very small $R$, Eq. (11) reduces to the point source solution (Eq. 7) at all energies.

\begin{figure}
\centering
\includegraphics*[width=0.3\textwidth,angle=270,clip]{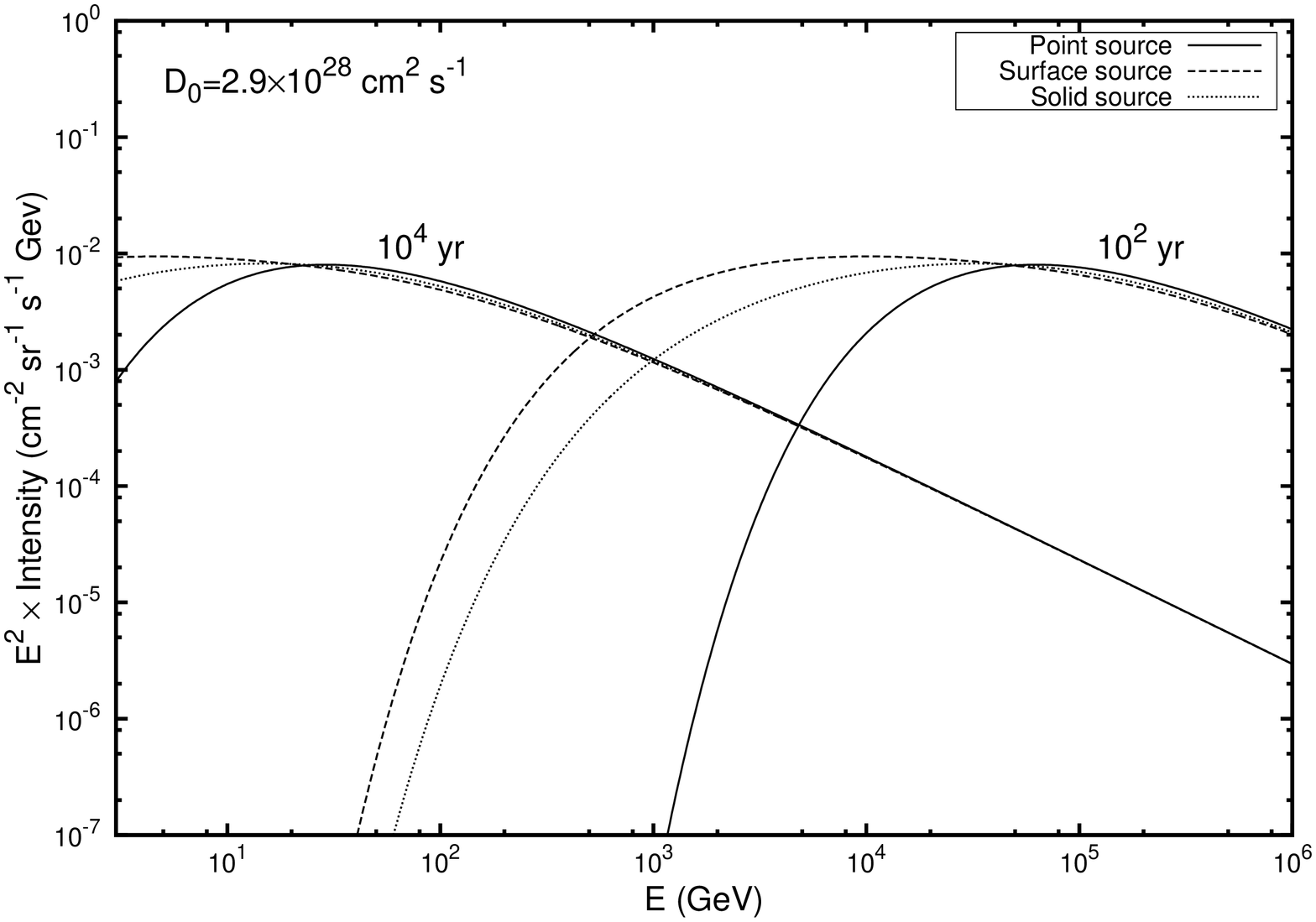}\\
\includegraphics*[width=0.3\textwidth,angle=270,clip]{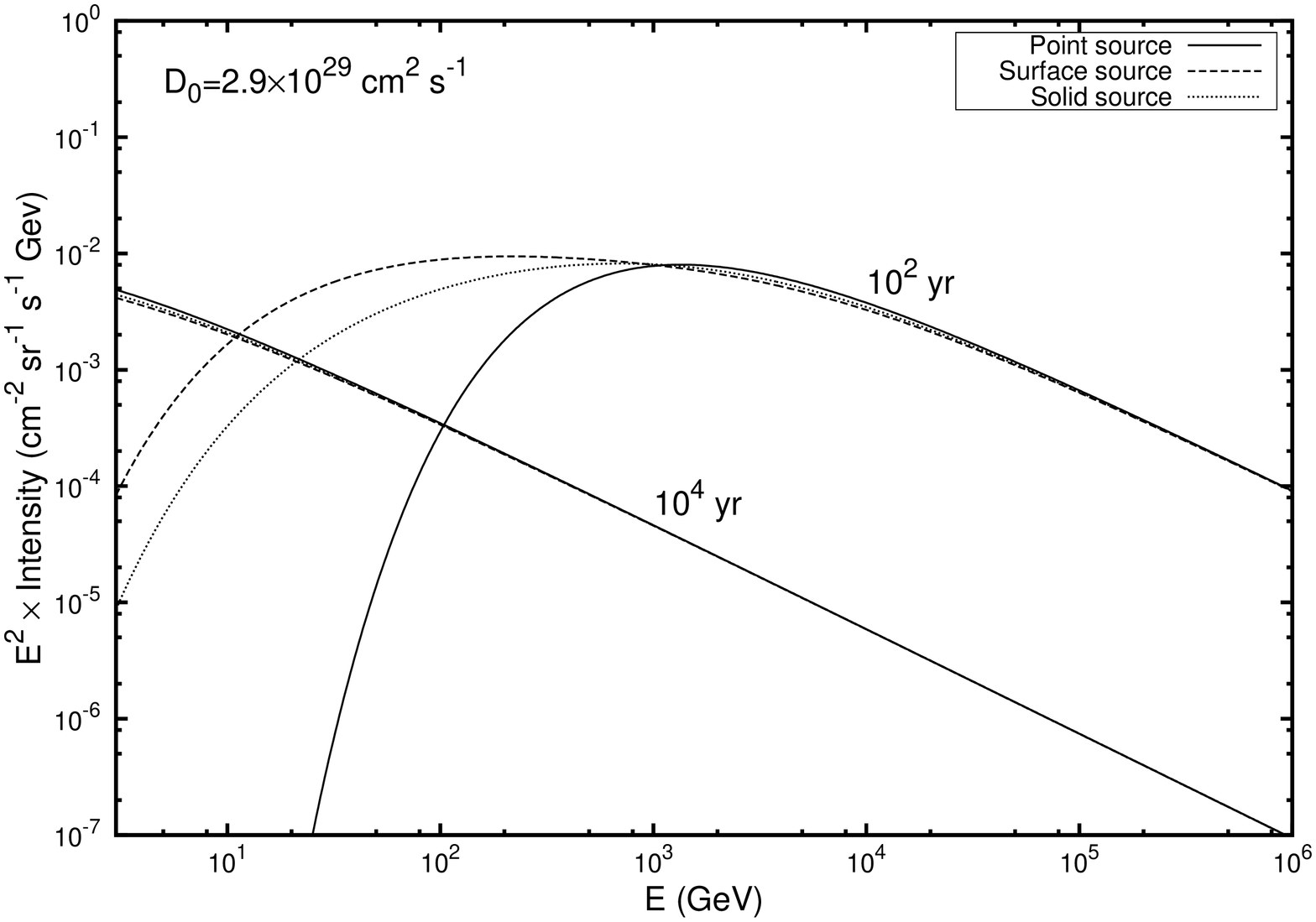}
\caption{\label {fig1} CR proton spectra from an SNR with distance $r_s=0.15$ kpc at different times $t=10^2$ yr and $10^4$ yr for different source models: point source (solid line), surface source (dashed line), solid source (dotted line). We consider an energy independent burst-like injection of CRs at $t_0=0$ and we take $\Gamma=2.0$, $\delta=0.6$ and $E_0=3$ GeV. Top: $D_0=2.9\times 10^{28}$ cm$^2$ s$^{-1}$. Bottom: $D_0=2.9\times 10^{29}$ cm$^2$ s$^{-1}$.}
\end{figure}
\begin{figure}
\centering
\includegraphics*[width=0.3\textwidth,angle=270,clip]{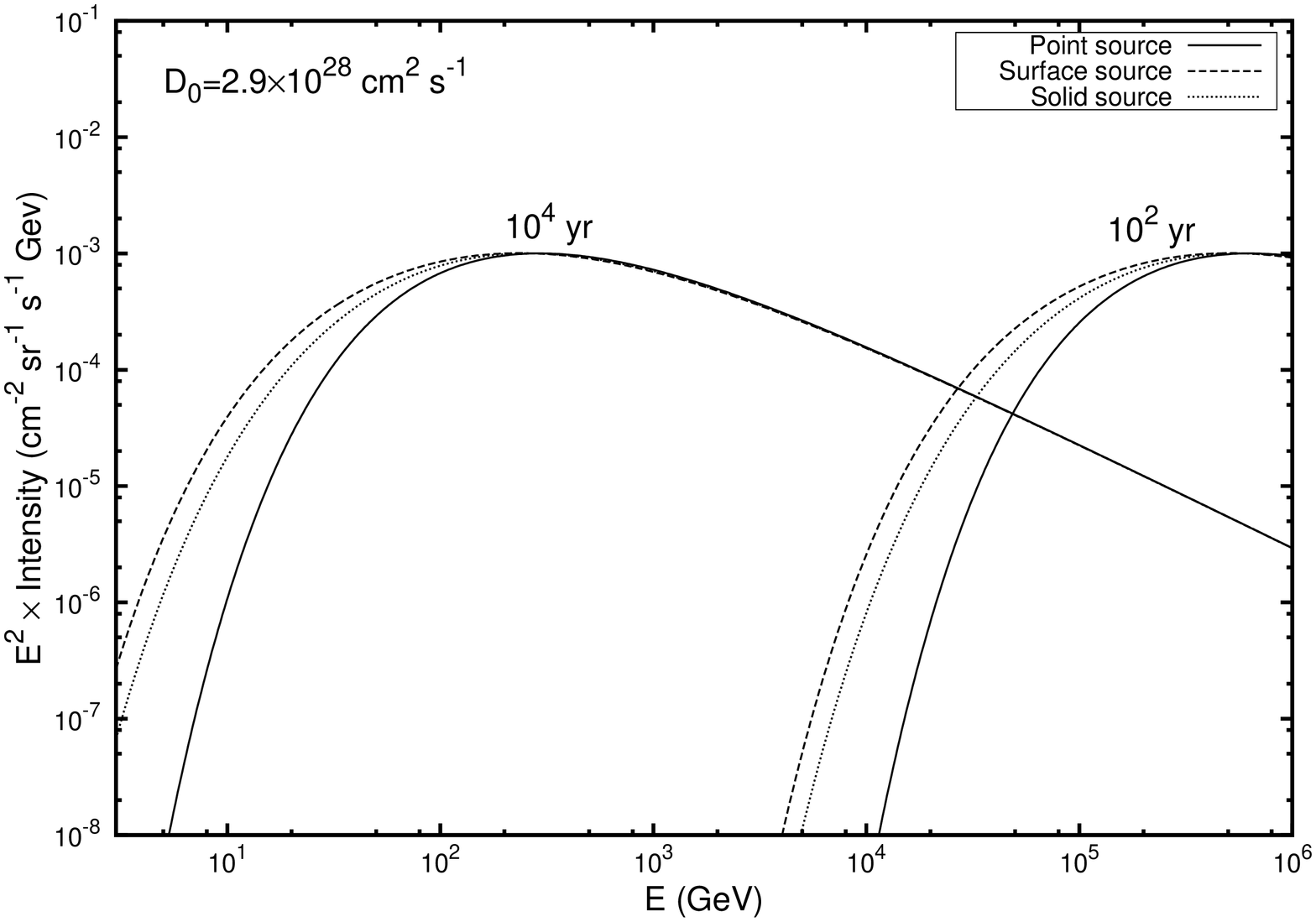}\\
\includegraphics*[width=0.3\textwidth,angle=270,clip]{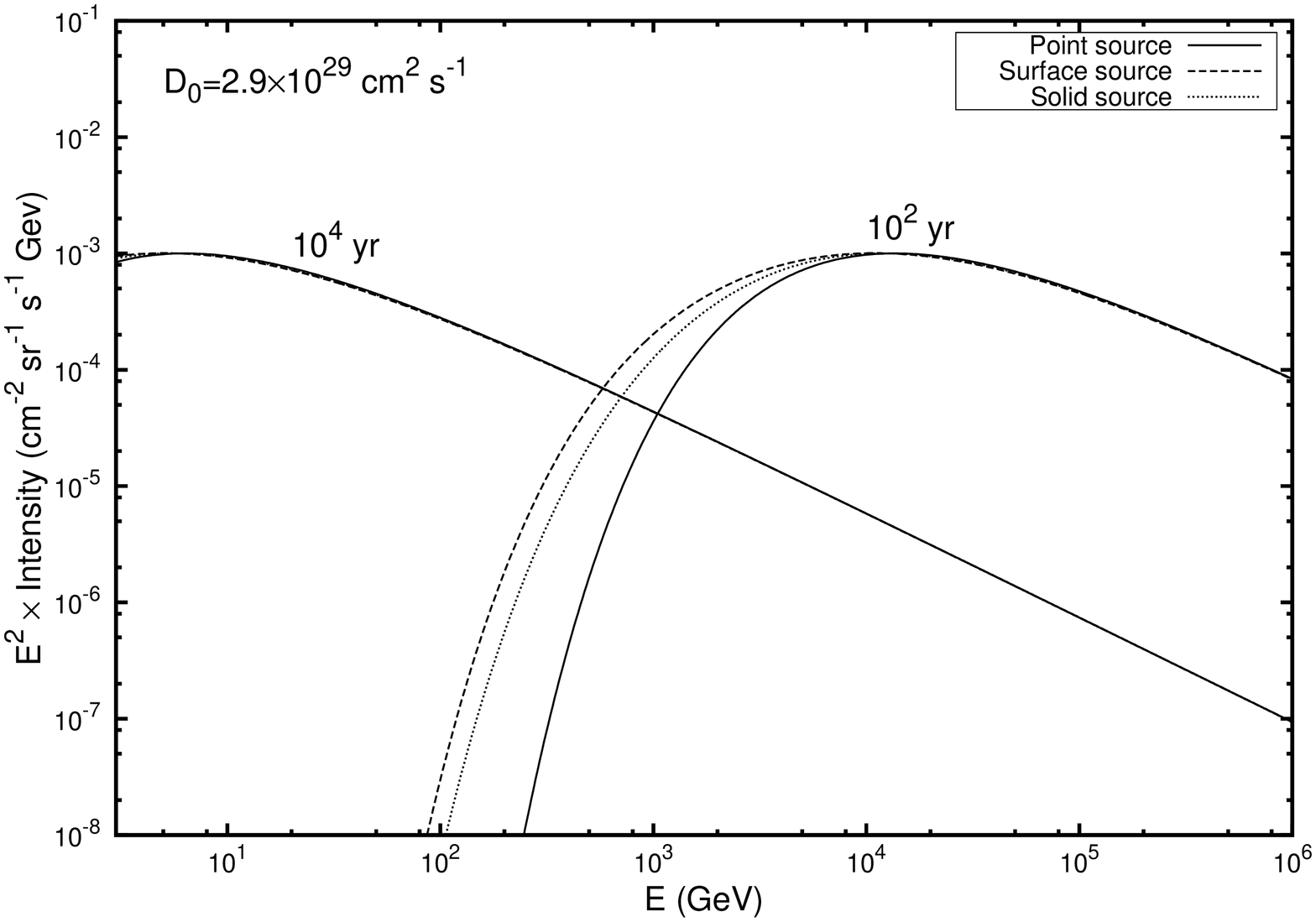}
\caption{\label {fig2} Same as in Fig. 1 but for $r_s=0.3$ kpc.}
\end{figure}
\begin{figure}
\centering
\includegraphics*[width=0.3\textwidth,angle=270,clip]{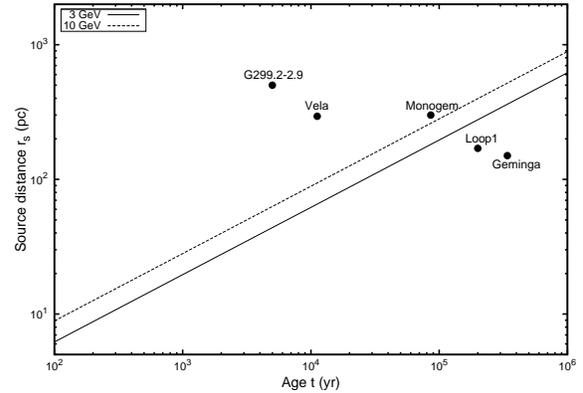}
\caption{\label {fig1} Source age $t$ versus distance $r_s$ plot for $E_{pt}=3$ GeV and $10$ GeV energies. The area below each line represents the $(r_s,t)$ parameters space where the point source represents a good approximation for energies $E> E_{pt}$. We assume $D_0=2.9\times 10^{28}$ cm $^2$ s$^{-1}$, $\delta=0.6$, $E_0=3$ GeV and $t_0=0$ for our calculation. The black dots represent the nearby known SNRs with distances $\leq 500$ pc (see Table 1).}
\end{figure}

\subsection{Spherical surface source}
If we assume that the CRs are distributed uniformly only on the surface of the SNR before they are released, the spatial source term in Eq. (9) can be written as,
\begin{equation}
q(r_0)=\delta(r_0-R)
\end{equation}
and the source spectrum as $q(E)=q_p(E)/A$, where $A=4\pi R^2$ denotes the total surface area of the SNR. The CR density in this case is then obtained as,
\begin{align}
N_p(E,t)=\frac{q_p(E)\,R}{r_sA\sqrt{\pi D(t-t_0)}}\mathrm{exp}\left[-\frac{\left(R^2+r_s^2\right)}{4D(t-t_0)}\right]\nonumber\\
\times\;\mathrm{sinh}\left(\frac{r_sR}{2D(t-t_0)}\right)
\end{align}
Here again, we can notice that Eq. (13) tends towards the point source solution for very small values of $R$. The spherical surface source model considered here closely represents the shell-type SNRs and it is probably more relevant than the solid source model for CR studies in our Galaxy. It is because according to the recent catalogue of Galactic SNRs, $78\%$ of the total known SNRs are of shell-type while the remaining $12\%$ and $4\%$ are of composite and plerion types respectively (Green 2009).

In Fig. 1 top panel, we compare the spectra obtained under the different source models for $t=10^2$ yr and $10^4$ yr. The calculation assumes $t_0=0$, the SNR distance as $r_s=0.15$ kpc and the diffusion constant as $D_0=2.9\times 10^{28}$ cm$^{2}$ s$^{-1}$. For our present illustration, we take the source spectral index as $\Gamma=2$ which is the value predicted by DSA theories inside SNRs. Later on, in section 5 when we apply our study to the nearby known SNRs, we will use values which are determined based on the observed CR data. In Fig. 1 top panel, we can see that for a given value of $t$, the point source solution (solid line) above some energy $E_{pt}$ agrees well with the results of the surface source (dashed line) and the solid source (dotted line) models, while below $E_{pt}$ the results are quite different. $E_{pt}$ is roughly the energy at which $r_{diff}=r_s$. We can check that for $E>E_{pt}$ for which $r_{diff}>r_s$, Eqs. (11) $\&$ (13) tend towards Eq. (8) which is the asymptotic solution of the point source approximation at high energies. The bottom panel shows the results for larger value of  diffusion constant $D_0=2.9\times 10^{29}$ cm$^{2}$ s$^{-1}$. The only difference between the two sets of results is that $E_{pt}$ is shifted towards lower values as $D_0$ increases. This shows that the point source becomes valid over a broader energy range as $D_0$ takes larger values. In order to understand the effect of the source distance, we show in Fig. 2 the results obtain for $r_s=0.3$ kpc by keeping all other parameters same as in Fig. 1. On comparing the results in Fig. 2 to those in Fig. 1, we can see that apart from the scaling down of the flux and the right shifting of $E_{pt}$ due to the increased source distance, the differences between the different source models also become smaller. This is simply the geometrical effect mentioned in section 1, i.e. as the source distance increases, the point source approximation becomes more valid.

These results can be understood as follows. The diffusion radius $r_{diff}=\sqrt{D(t-t_0)}$, which is the effective distance from the SNR travelled by CRs due to diffusive propagation, is a strong function of $D_0$ and $t$. The larger the values of $D_0$ and/or $t$, the larger is $r_{diff}$ and the energy $E_{pt}$ which satisfy the condition $r_{diff}=r_s$ decreases. Similarly, for larger source distances $r_s$, we can understand that $E_{pt}$ shifts towards higher values. These results show that the point source can remain a valid approximation even for the nearby sources as long as the particles satisfy the condition $r_{diff}\gg(R,r_s)$. For $r_s=0.15$ kpc, $D_0=2.9\times 10^{28}$ cm$^{2}$ s$^{-1}$ and $t_0=0$ (Fig.1 top panel), we obtain $E_{pt}=1.2\times 10^5$ GeV and $57$ GeV for $t=10^2$ yr and $10^4$ yr respectively. The corresponding values for $r_s=0.3$ kpc (Fig. 2 top panel) are found to be $1.2\times 10^6$ GeV and $575$ GeV respectively. For a given age or distance, the closer or the older the source is, the lower is the $E_{pt}$. This is shown in Fig. 3 for $E_{pt}=3$ GeV (solid line) and $10$ GeV (dashed line). The area below each line represents the parameters space in $(r_s,t)$ where the point source approximation works as a good approximation for all energies above the given $E_{pt}$. In the same figure, the black dots represent the nearby known SNRs with distances $\leq 500$ pc (see Table 1). We can see that only the Loop1 and the Geminga remnants lie below both the lines while the Monogem lie just above the $10$ GeV line. The other two SNRs, Vela and G299.9-2.9 are located well above the lines. It is worth mentioning that CR spectrum below $\sim (1-10)$ GeV are very much likely to be modified by the solar modulation and hence, only those above this energy region are reliable for estimates of Galactic CR properties. Therefore, as far as the CRs of our interests are concerned, the point source approximation looks valid only for the Loop1, Geminga and the Monogem among the nearest SNRs, while for the others it looks important to take their sizes into account in the calculations. It should be noted that for distant SNRs like SN185 for which the distances $r_s\gg R$, the point source will always remain a good approximation independent of their ages. We will show this in detail in section 5.

The results shown in Figs. 1 and 2 are obtained for the burst-like injection of particles. Let us now investigate the case of continuous injection of particles. For continuous injection for a finite time interval from $0$ to $T$ with the injection rate $q_c(E)$, the solution of Eq. (1) can be obtained by integrating Eq. (5) over the injection time as,
\begin{align}
N_p(E,t)=\int_0^{t_f}dt^\prime\int^{\infty}_{-\infty}d\textbf{r}^\prime\frac{q_c(E)}{8\left[\pi D(t-t^\prime)\right]^{3/2}}\mathrm{exp}\left[\frac{-{\textbf{r}^\prime}^2}{4D(t-t^\prime)}\right]\nonumber\\
\times\;q(\textbf{r}^\prime)
\end{align}
where $t_f=\mathrm{min}[t,T]$ and $q_c(E)=q(E)/T$, $q(E)$ representing the source spectrum in the burst-like injection case which for a point source is given by $q(E)=q_p(E)$. For a point source positioned at $\textbf{r}_s$, Eq. (14) becomes, 
\begin{equation}
N_p(E,t)=\frac{q_p(E)}{4\pi r_sDT}\lbrace\mathrm{erf}\left(\sqrt{x_2}\right)-\mathrm{erf}\left(\sqrt{x_1}\right)\rbrace
\end{equation}
where,
\begin{equation*}
x_2=\frac{r_s^2}{4D(t-t_f)}\quad\mathrm{and}\quad x_1=\frac{r_s^2}{4Dt}
\end{equation*}

For $t<T$, $t_f = t$ which implies $x_2 =\infty$. Then, using the property of the error function, $\mathrm{erf}(\sqrt{x_2}) = 1$ for $x_2 = \infty$, Eq. (15) in this case becomes,
\begin{equation}
N_p(E,t)=\frac{q_p(E)}{4\pi r_sDT}\lbrace 1-\mathrm{erf}\left(\sqrt{x_1}\right)\rbrace
\end{equation}
For high-energy particles for which the diffusion radius $r_{diff}\gg r_s$, $x_1\rightarrow 0$, and because $\mathrm{erf}(\sqrt{x_1})\rightarrow 0$ for $x_1\rightarrow 0$, the spectrum given by Eq. (16) follows a power law of the form $N_p(E)\propto E^{-(\Gamma+\delta)}$ which is flatter than the spectrum we obtain in the burst-like injection model. A detailed discussion on this topic can also be found in Aharonian $\&$ Atoyan 1996 in the study of CR spectrum in the vicinity of the sources.

\begin{figure}
\centering
\includegraphics*[width=0.3\textwidth,angle=270,clip]{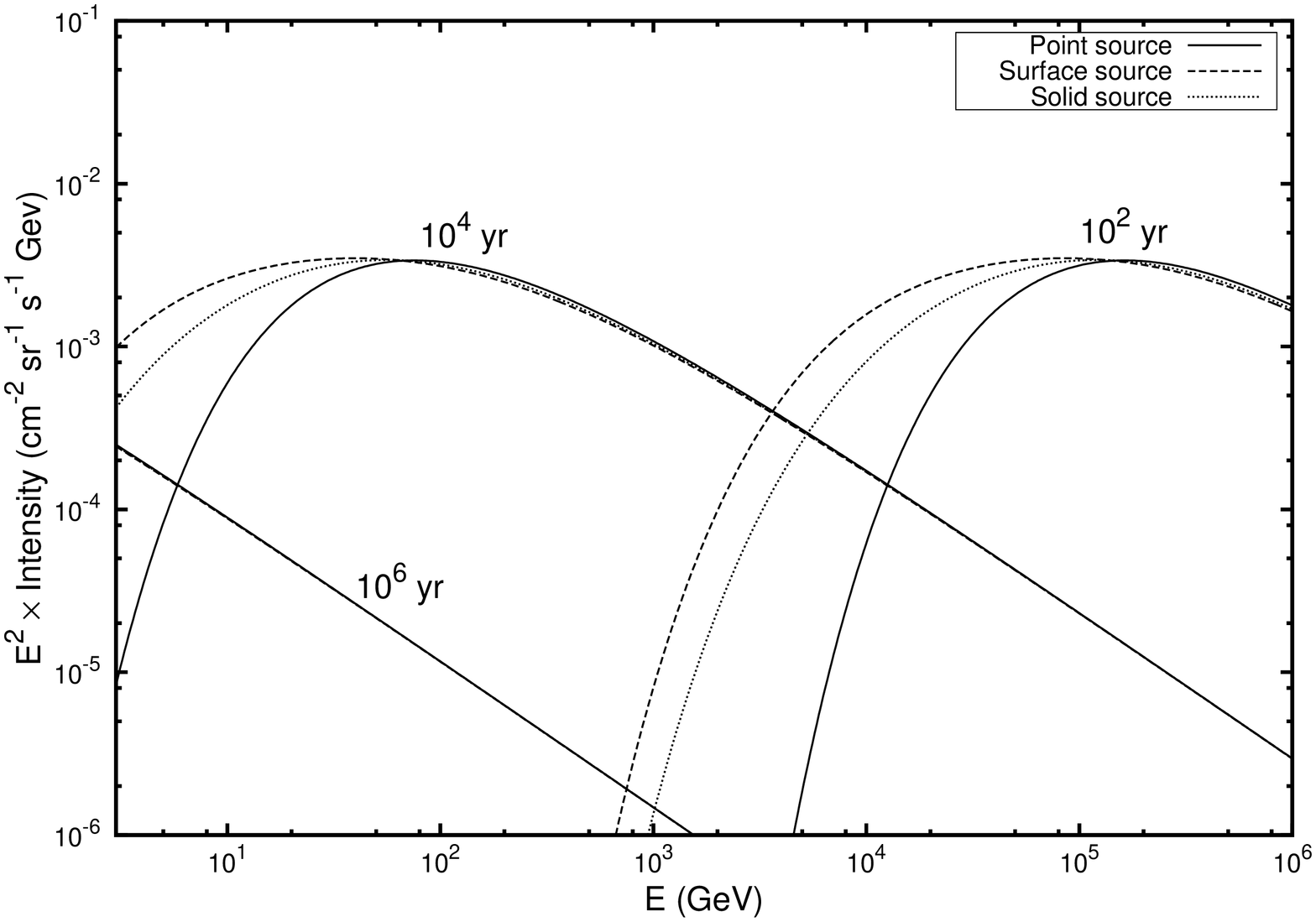}\\
\includegraphics*[width=0.3\textwidth,angle=270,clip]{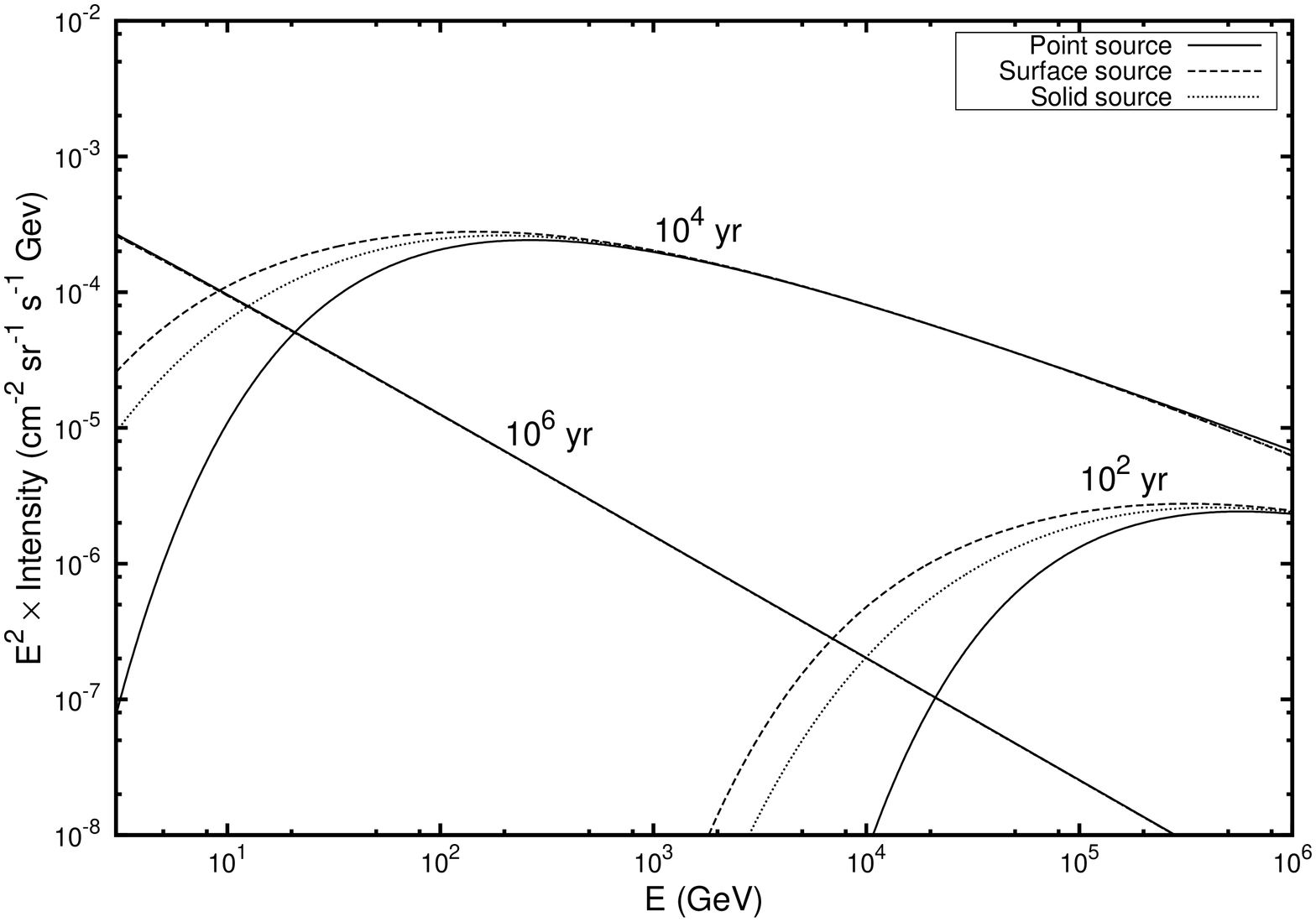}
\caption{\label {fig1} CR proton spectra for different source models from an SNR at $r=0.2$ kpc at different times $t=(10^2, 10^4, 10^6)$ yr. Top: Burst-like injection. Bottom: Continuous injection with  $T=10^5$ yr. Other model parameters are the same as in Fig. 1 top panel.}
\end{figure}

For $t>T$, $t_f = T$ and $x_2 = r_s^2/4D(t-T)$. For particles with large $r_{diff}$ for which $x_1\rightarrow 0$, we can safely write $x_2\ll 1$ as $(t-T)<t$. Then, using the property $\mathrm{erf}(\sqrt{x_2}) \approx  2\sqrt{x_2/\pi}$ for $\sqrt{x_2}\ll 1$, the particle spectrum (Eq. 15) in this case reduces to
\begin{equation}
N_p(E,t)\approx\frac{q_p(E)}{4(\pi D)^{3/2}T\sqrt{t-T}}
\end{equation}
The spectral shape of Eq. (17) follows $N_p(E)\propto E^{-\left(\Gamma+\frac{3}{2}\delta\right)}$ which is the same as in the case of the burst-like injection of particles discussed earlier (Eq. 8).

Similarly, using Eq. (14) we also obtain our results for the case of the solid and the surface source models by taking into account their proper source terms given by Eqs. (10) and (12) respectively. The results are plotted in Fig. 4 (bottom panel) for $t=(10^2, 10^4, 10^6)$ yr along with the results obtained under the burst-like injection model (top panel) for comparison. The calculations in Fig. 4 assume $r_s=0.2$ kpc, $D_0=2.9\times 10^{28}$ cm$^2$s$^{-1}$, $\Gamma=2.0$ and $T=10^5$ yr. We can see that at all $t$'s, the effect of assuming different source models are similar in both the types of injection. As discussed above, we can also see that for $t<T$ the spectra in the case of continuous injection are flatter than those in the burst-like injection case while at $t\gg T$, they exactly follow the same behavior as shown by the results at $10^6$ yr.

A short conclusion that we can draw at this stage of our study is that for very old sources ($t\gg T$), the effect of choosing different source geometry or different particle injection model is negligible on the CR spectrum. Therefore, the widely adopted burst-like point source approximation remains a good approximation for very old nearby sources  at all the energies. However, for young nearby sources, the spectrum at high energies strongly depends on type of the particle injection model and at lower energies, it starts depending on the physical size and the geometry of the source irrespective of the type of the injection model unless the source is really closed to the Earth i.e., only a few pc away as shown in Fig.3.
 
\section{High energy electron spectrum from an SNR}
The diffusive propagation of high energy electrons in the Galaxy can be described by the following transport equation,
\begin{equation}
\nabla\cdot(D\nabla N_e)+\frac{\partial}{\partial E}\lbrace b(E) N_e\rbrace+Q_e=\frac{\partial N_e}{\partial t}
\end{equation} 
where $N_e(E,t)$ is the density of electrons with kinetic energy $E$, $b(E)=-dE/dt$ is the energy loss rate and $Q_e(\textbf{r},E,t)$ denotes the electron injection rate into the ISM. The Green function of Eq. (18) can be obtained as given below (see e.g., Ginzburg $\&$ Syrovatskii 1964, Gratton 1972),
\begin{align}
G_e(\textbf{r},\textbf{r}^\prime,E,E^\prime,t,t^\prime)=\frac{1}{8\left[\pi f(E,E^\prime) \right]^{3/2}b(E)}\mathrm{exp}\left[\frac{-(\textbf{r}^\prime -\textbf{r})^2}{4f(E,E^\prime)}\right]\nonumber\\
\times\;\delta\left[t^\prime-t+g(E,E^\prime)\right]
\end{align}
where,
\begin{equation*}
f(E,E^\prime)=\int_E^{E^\prime} \frac{D(u)}{b(u)}du\quad \mathrm{and}\quad g(E,E^\prime)=\int_E^{E^\prime}\frac{1}{b(u)}du
\end{equation*}
For our present study, we assume that the energy loss of the electrons are due to synchrotron and inverse compton interactions which are true mostly for energies $E\gtrsim 10$ GeV. We take,
\begin{equation}
b(E)=aE^2
\end{equation}
where $a=1.01\times 10^{-16}(w_{ph}+w_B)$ GeV s$^{-1}$ and, $w_{ph}$ and $w_B$ represent the energy densities in eV cm$^{-3}$ for the background photons and the magnetic field respectively. Eq. (20) assumes that the inverse compton scattering of the background photons occur in the Thompson regime.

The general solution of Eq. (18) is given by,
\begin{align}
N_e(\textbf{r},E,t)=\int^{\infty}_{-\infty}d\textbf{r}^\prime\int_E^{\infty}dE^\prime\int^t_{-\infty}dt^\prime G_e(\textbf{r},\textbf{r}^\prime,E,E^\prime,t,t^\prime)\nonumber\\
\times\;Q_e(\textbf{r}^\prime,E,t^\prime)
\end{align}
For an energy independent burst-like injection of electrons at time $t_0$, we take the source term as $Q_e(\textbf{r}^\prime,E^\prime,t^\prime)=q(\textbf{r}^\prime)q(E^\prime)\delta(t^\prime-t_0)$ where $q(E^\prime)\propto E^{\prime -\Gamma}$ denotes  the source spectrum. Now, setting $\textbf{r}=0$ as we did for the protons in section 2 and performing the integrals over $E^\prime$ and $t^\prime$, Eq. (21) becomes,
\begin{align}
N_e(E,t)=\frac{q(E)}{8(\pi C)^{3/2}}\left(1-\frac{E}{E_t}\right)^{\Gamma-2}\int_\infty^{-\infty}d\textbf{r}^\prime\nonumber\\
\times\;\mathrm{exp}\left(\frac{-{\textbf{r}^\prime}^2}{4C}\right)q(\textbf{r}^\prime)
\end{align}
where $E_t=1/(a(t-t_0))$ is the energy at which the energy loss time is equal to $(t-t_0)$,
\begin{equation}
C=\frac{D(E)}{a(1-\delta)E}\left[1-\left(1-\frac{E}{E_t}\right)^{1-\delta}\right]
\end{equation}
and $\delta$ is the index of the diffusion coefficient. Eq. (22) is valid for electrons with energies $E<E_t$. For $E>E_t$, $N_e=0$.

\begin{figure}
\centering
\includegraphics*[width=0.3\textwidth,angle=270,clip]{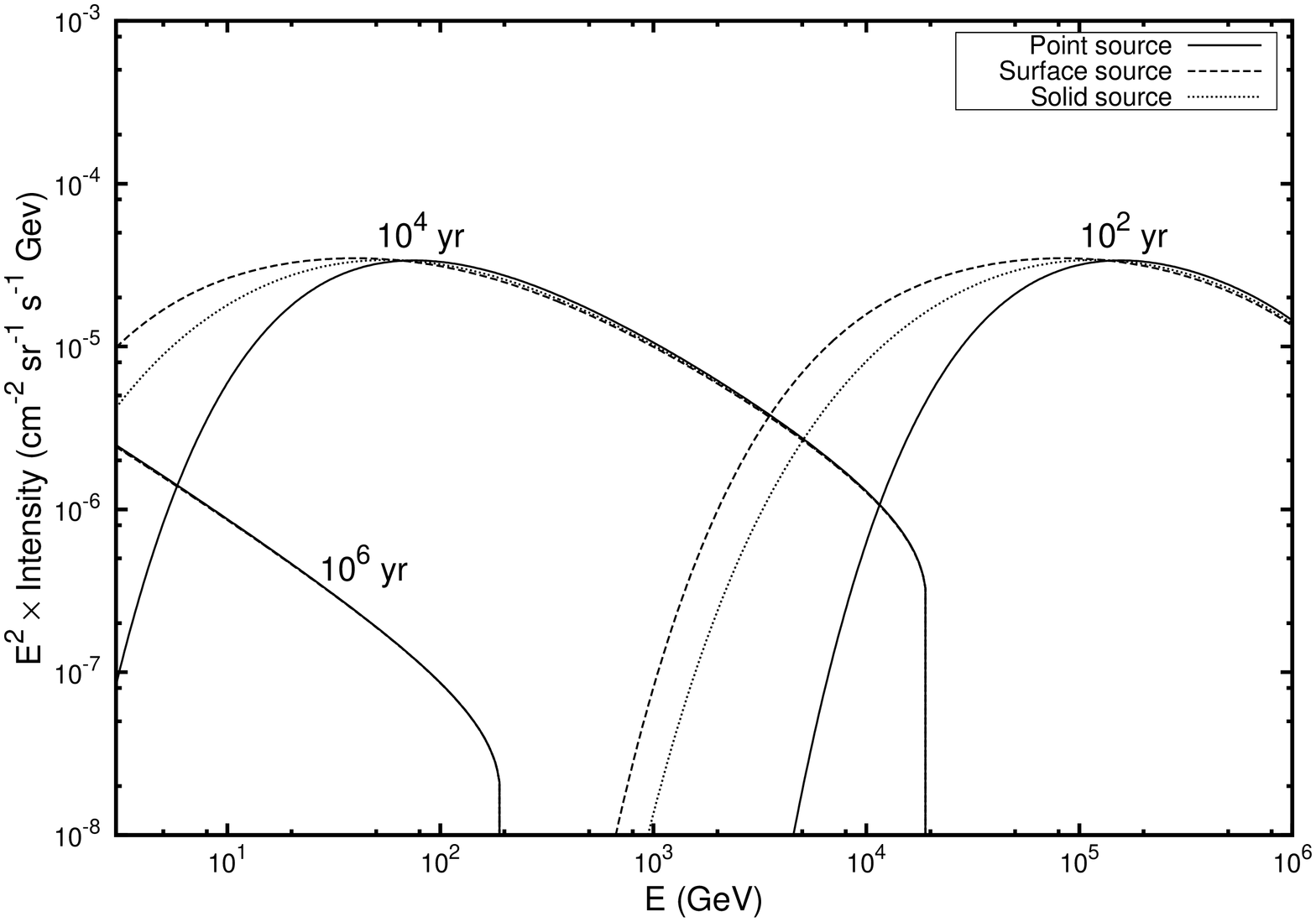}\\
\includegraphics*[width=0.3\textwidth,angle=270,clip]{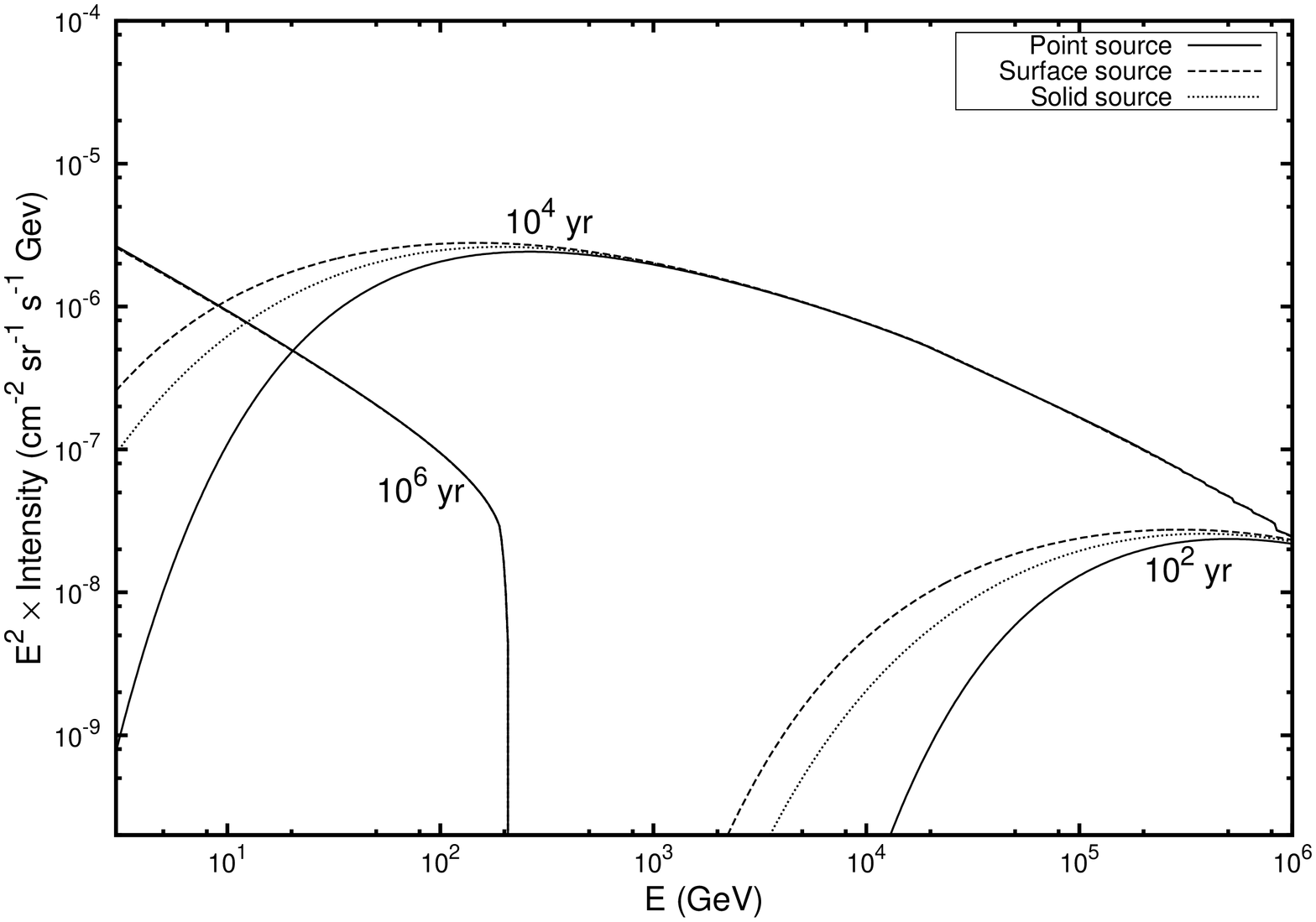}
\caption{\label {fig1} CR electron spectra for different source models from an SNR at $r_s= 0.2$ kpc. We assume $w_{MBR}= 0.25$ eV cm$^{-3}$, $w_{op} = 0.5$ eV cm$^{-3}$ and the ISM magnetic field as $6\mu$G. All other model parameters are the same as in Fig. 4. Top: Burst-like injection. Bottom: Continuous injection.}
\end{figure}

\subsection{Point source approximation}
For a point source described by Eq. (6) located at a distance $\textbf{r}_s$, the electron spectrum at time $t$ can be obtained from Eq. (22) as given below,
\begin{equation}
N_e(E,t)=\frac{q_e(E)}{8(\pi C)^{3/2}}\left(1-\frac{E}{E_t}\right)^{\Gamma-2}\mathrm{exp}\left(\frac{-{\textbf{r}_s}^2}{4C}\right)
\end{equation}
where $q_e(E)=k_e E^{-\Gamma}$ is the source spectrum for the point source. From Eq. (23), we can see that in the energy region $E\ll E_t$ where the effect of the energy loss is less important, $C\rightarrow D(t-t_0)$ and Eq. (24) tends towards the point source solution for CR protons (Eq. 7). Therefore, high energy electrons whose $r_{diff}\gg r_s$ and $E\ll E_t$ have spectrum which follows $N_e(E)\propto E^{-\left(\Gamma+\frac{3}{2}\delta\right)}$ which is similar to the asymptotic solution of high energy protons (Eq. 8).

\subsection{Spherical solid source}
Following exactly the same procedure as for the protons described in the previous section, we obtain the electron spectrum for the spherical solid source as,
\begin{align}
N_e(E,t)=\frac{q_e(E)}{r_sV\sqrt{\pi C}}\left(1-\frac{E}{E_t}\right)^{\Gamma-2}\mathrm{exp}\left(\frac{-r_s^2}{4C}\right)\nonumber\\
\times\;\int^{R}_0r_0\,\mathrm{exp}\left(\frac{-r_0^2}{4C}\right)\mathrm{sinh}\left(\frac{r_sr_0}{2C}\right)dr_0
\end{align}

\subsection{Spherical surface source}
We also obtain the solution for the spherical surface source as given below,
\begin{align}
N_e(E,t)=\frac{q_e(E)\,R}{r_sA\sqrt{\pi C}}\left(1-\frac{E}{E_t}\right)^{\Gamma-2}\mathrm{exp}\left[-\frac{\left(R^2+r_s^2\right)}{4C}\right]\nonumber\\\times\;\mathrm{sinh}\left(\frac{r_sR}{2C}\right)
\end{align}
It is easy to check that for very small values of $R$, the solutions for the solid source (Eq. 25) and the surface source (Eq. 26) models reduce to the point source solution (Eq. 24).

The solutions we have obtained above are based on  burst-like injection of electrons from the SNR. For the case of continuous injection, the solutions are given by,
\begin{align}
N_e(E,t)=\int_0^{t_f}dt^\prime\frac{q_c(E)}{8(\pi C)^{3/2}}\left(1-\frac{E}{E_t^\prime}\right)^{\Gamma-2}\nonumber\\
\times\;\int_\infty^{-\infty}d\textbf{r}^\prime \mathrm{exp}\left(\frac{-{\textbf{r}^\prime}^2}{4C}\right)q(\textbf{r}^\prime)
\end{align}
where $t_f$ and $q_c(E)$, they bear the same definitions as defined in the case of protons and $E_t^\prime=1/(a(t-t^\prime))$. For a point source at $\textbf{r}_s$, Eq. (27) becomes,
\begin{align}
N_e(E,t)=\int_0^{t_f}dt^\prime\frac{q_e(E)}{8(\pi C)^{3/2}T}\left(1-\frac{E}{E_t^\prime}\right)^{\Gamma-2}\nonumber\\
\times\;\mathrm{exp}\left(\frac{-{\textbf{r}_s}^2}{4C}\right)
\end{align}
where $q_e(E)$ is the source spectrum we assumed in the case of burst-like injection (section 3.1). In Eqs. (27) and (28), $C$ is given by Eq. (23) but with $E_t$ replaced by $E_t^\prime$ in this case. Here again, we can check that for energies $E\ll E_t^\prime$, $C\rightarrow D(t-t^\prime)$ and Eq. (28) reduces to a solution similar to that of the CR protons (Eq. 15). Therefore, the same discussions we presented in the previous section for the protons under the continuous injection model also apply to the electrons. At time $t<T$, the spectrum of high energy electrons with $E\ll E_t^\prime$ and whose diffusion radii $r_{diff}\gg r_s$ follow $N_e(E)\propto E^{-(\Gamma+\delta)}$ and at $t>T$, they follow $N_e(E)\propto E^{-\left(\Gamma+\frac{3}{2}\delta\right)}$ which is similar to the result obtained in the burst-like injection model (section 3.1). More discussions on the different types of electron spectra generated by a CR source under different particle injection models can also be found in Atoyan et al. 1995.

Using Eq. (27), we also calculate the spectra for the solid and the surface sources under the continuous injection model. The results are plotted in Fig. 5 (bottom panel) for $t=(10^2, 10^4, 10^6)$ yr. The top panel shows the results for the case of burst-like injection. The calculations assume $\Gamma=2.0$, $\delta=0.6$, $D_0=2.9\times 10^{28}$ cm$^2$ s$^{-1}$, $r_s=0.2$ kpc, $T=10^5$ yr and $t_0=0$. The magnetic field in the ISM is taken as $6\mu$G (Beck 2001). The total energy density of the background photon field is assumed to be $w_{ph}=w_{MBR}+w_{op}$, where $w_{MBR}=0.25$ eV cm$^{-3}$ is the energy density of the microwave background and $w_{op}=0.6$ eV cm$^{-3}$ that of the ultraviolet-NIR-optical radiation field. The latter is taken from the estimates given in Shibata et al. 2011 for the galactocentric distance of $8.5$ kpc which they obtain using the data provided by GALPROP (Porter et al. 2008). On comparing Fig. 5 top and bottom panels, we can notice that at $t<T$ apart from the difference in the slope of the spectra, there are sharp spectral breaks present in the case of burst-like injection. These are due to the effect of fast energy loss rate for high energy electrons. Electrons with energy $E>E_t$ are lost before reaching the Earth. We also notice that the differences between the different source models below $E_{pt}$ are similar in both the types of injection model. At very late times $t\gg T$, the spectra becomes independent of the injection model or of the source model and they exhibit the same shapes with breaks at $E=E_t$.

On comparing the results of electrons (Fig. 5) to those of the protons (Fig. 4), we can see that except for the presence of spectral breaks in the case of electrons, the results are quite similar in all other respects. Even the differences between the results obtained for different source models are similar. Therefore, the overall conclusions on the validity of the point source approximation that we had drawn earlier for the protons also apply to the electrons.

\section{Energy dependent CR escape from SNRs}
So far, we have only considered a simple model of CR escape from the SNRs where CRs of all energies are assumed to escape at the same time independent of energy. However, detailed theoretical studies suggest that their escape mechanism can be more complex which may be  strongly related with the acceleration process itself and depends on the shock dynamics as well as on the particle energies and their back reaction on the shocks (see e.g. Malkov $\&$ Drury 2001, Ptuskin $\&$ Zirakashvili 2005). 

As already mentioned in section 1, under DSA theory CRs are assumed to be confined by the magnetic turbulence generated by the CRs themselves. They cannot escape the remnant as long as their upstream diffusion length normally defined as $l_{diff}=D_s(E)/u_s$ is less than the escape length from the shock front which is usually taken as $l_{esc}=\xi R_s$, where $u_s$ and $R_s$ denote the shock velocity and the shock radius respectively and the constant $\xi\sim (0.01-0.1)$ (Ptuskin $\&$ Zirakashvili 2005 and references therein). In the Bohm diffusion limit where the maximum confinement is acheivable, the upstream diffusion coefficient depends on the particle energy and the upstream magnetic field $B_s$ as $D_s(E)\propto E/B_s$. Under this condition, the escape energy $E_{esc}$ follows,
\begin{equation}
E_{esc}\propto B_sR_su_s
\end{equation}

There are strong theoretical arguments which suggest that CRs might amplify the magnetic fields near the shock surface (see e.g. Caprioli et al. 2009). This idea is also supported experimentally by the recent observations of thin X-ray filaments inside several SNRs, which are most likely synchrotron emissions of high energy electrons in the presence of strong magnetic fields of the order of $\sim (100-1000)\mu$G (V\"{o}lk et al. 2005). Taking such possible amplification into account, we can assume that the magnetic field scales with the shock velocity as $B_s\propto u_s^d$, with the index $d$ representing the degree of amplification. Some studies suggest that $d$ can reach values as high as $1.5$ (Bell 2004).

One reasonable assumption of DSA theory is that CRs do not escape during the free expansion phase of the SNR evolution. It is because shock waves traveling at some constant velocity can always overtake particles undergoing diffusive motion (Drury 2011). However, during the Sedov phase when the shock velocity decreases with the age $t$ as $u_s\propto t^{-0.6}$ and the shock radius increases as $R_s\propto t^{0.4}$, some of the high energy CRs can start escaping because of their relatively larger diffusion length $(l_{diff}>l_{esc})$. Therefore, under the Sedov scaling, the escape energy at any stage during the evolution can be obtained using Eq. (29) as,
\begin{equation}
E_{esc}\propto t^{-(0.2+0.6d)}
\end{equation}
This gives,
\begin{equation}
E_{esc}\propto\begin{cases}t^{-0.2}, &\text{for $d=0$}\\
t^{-1.1}, &\text{for $d=1.5$}\end{cases}
\end{equation}

In deriving Eq. (31), we assume that $D_s(E)$ scales linearly with $E$. But, the exact dependence is still not well understood and depends on some poorly known yet important quantities like the spectral distribution of the self-excited turbulence waves, their dissipation rate and their CR scattering efficiencies. Moreover, magnetic field amplification and the dynamical reaction of the accelerated particles on the shock structure are also not fully understood. Due to these uncertainties, a simple but reasonable approach which is commonly followed is to parameterize the escape energy as given below (Gabici et al. 2009, Ohira et al. 2011),
\begin{equation}
E_{esc}=E_{max} \left(\frac{t}{t_{sed}}\right)^{-\alpha}
\end{equation}
where $E_{max}$ is the maximum CR energy and $t_{sed}$ denotes the start of the Sedov phase. We assume $E_{max}=10^6$ GeV ($=1$ PeV) and $t_{sed}=500$ yr for our study. Eq. (32) assumes that the escape of the highest energy particles start at the onset of the sedov phase itself. For detailed studies of particle escape from SNRs, see e.g. Ptuskin $\&$ Zirakashvili 2005, Caprioli et al. 2009, Caprioli et al. 2010. Using Eq. (32), we can calculate the escape time $t_{esc}$ as a function of energy as,
\begin{equation}
t_{esc}(E)=t_{sed}\left(\frac{E}{E_{max}}\right)^{-1/\alpha}
\end{equation}

At some later stage of the SNR evolution when the shock slows down and does not efficiently accelerate the CRs, the turbulence level in the vicinity of the shock goes down and no particles can remain confined effectively within the remnant. At this stage, we can assume that all the CRs escape into the ISM. As previously mentioned, for an ISM density of 1 H cm$^{-3}$, this happens at around $10^5$ yr after the supernova explosion (Berezhko $\&$ V\"olk 2000). Taking this into account, the CR escape time for our study is taken as,
\begin{equation}
T_{esc}(E)=\mathrm{min}\left[t_{esc}(E),10^5 \mathrm{yr}\right]
\end{equation}

Using the Sedov relation between the shock radius and the SNR age, we can also calculate the escape radius $R_{esc}$ which we define as the radius of the SNR at the time when CRs of energy $E$ escape as follows,
\begin{equation}
R_{esc}(E)=2.5v_0\;t_{sed}\left[\left(\frac{T_{esc}}{t_{sed}}\right)^{0.4}-0.6\right]
\end{equation}
In Eq. (35), $v_0$ represents the initial shock velocity, i.e the velocity at $t=t_{sed}$ which we take as $10^9$ cm/s for our study.
\begin{figure}
\centering
\includegraphics*[width=0.3\textwidth,angle=270,clip]{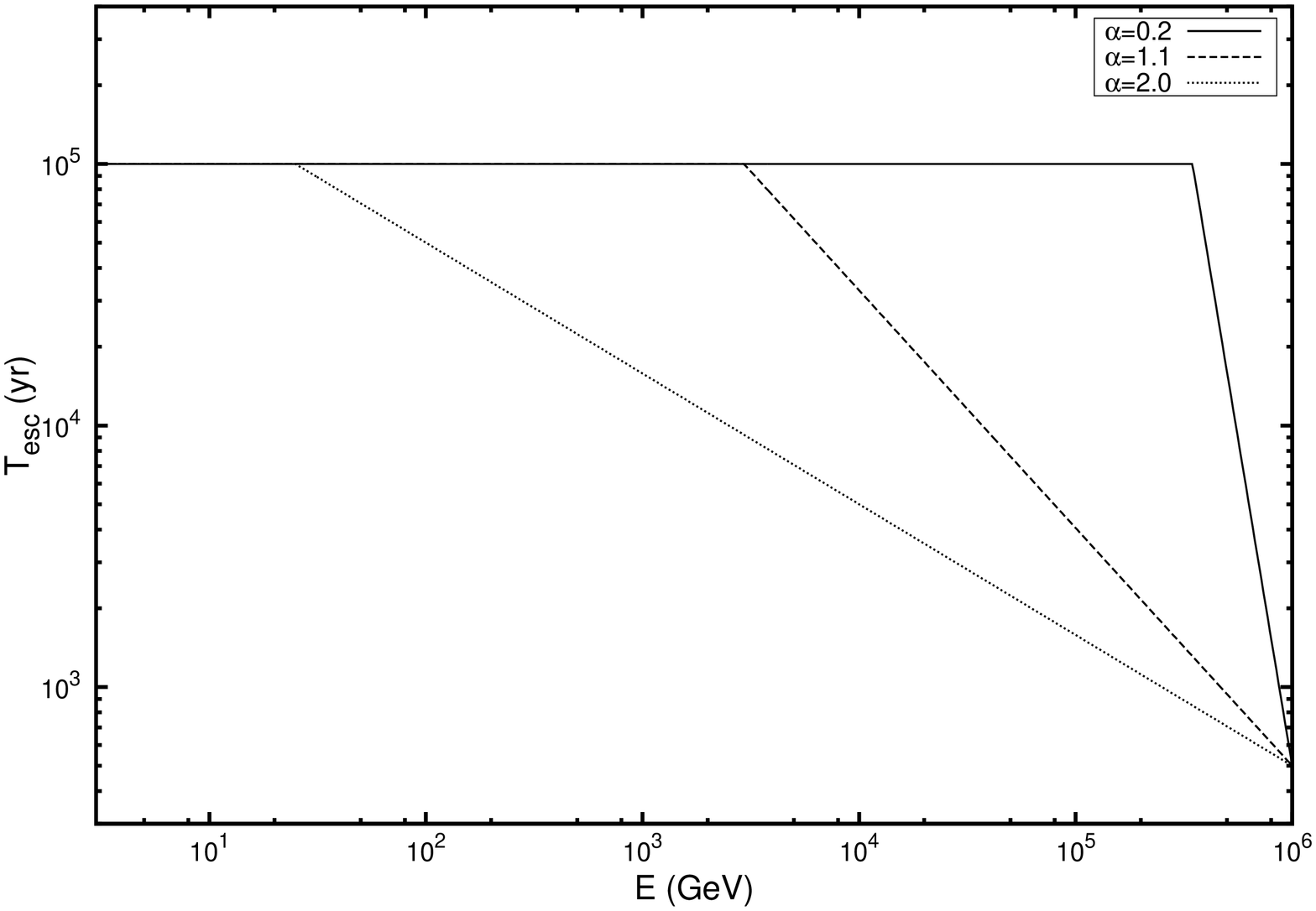}\\
\includegraphics*[width=0.3\textwidth,angle=270,clip]{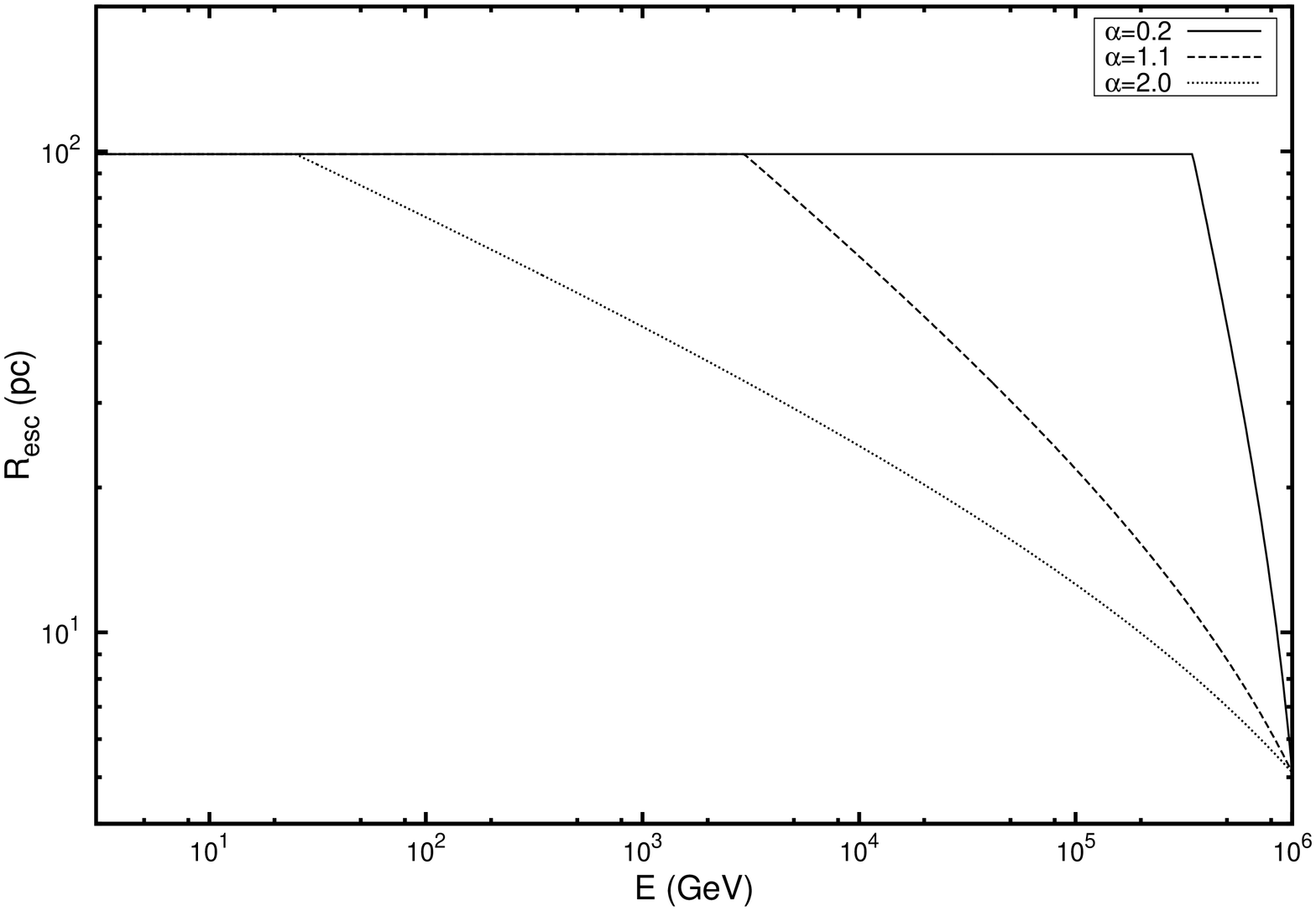}
\caption{\label {fig1} CR escape time (top) and escape radius (bottom) for an SNR under the energy dependent escape model for different values of $\alpha$: 0.2 (solid line), 1.1 (dashed line), 2.0 (dotted line). We assume $E_{max}=10^6$ GeV and $t_{sed}=500$ yr.}
\end{figure}

Eq. (34) is plotted in Fig. 6 (top panel) where different lines correspond to different values of $\alpha$: solid (0.2),  dashed (1.1) and dotted (2.0). The plots show that even for the fixed values of $E_{max}$ and $t_{sed}$,  the energy dependence of $T_{esc}(E)$ strongly depends on the value of $\alpha$. For $\alpha=0.2$, except for particles with energies greater than $3.5\times 10^5$ GeV all the particles remain confined till the end of the SNR evolution. As the value of $\alpha$ increases, lower energy particles start escaping at relatively early stages. For $\alpha=1.1$ and $2.0$, only particles with energies up to $3\times 10^3$ GeV and $25$ GeV respectively are confined till the end of the evolution. The bottom panel shows the corresponding values of $R_{esc}(E)$ calculated using Eq. (35). For the assumed value of $v_0$, CRs escape starts when the remnant expands to a radius of $\sim 5$ pc and continues until it expands up to $\sim 100$ pc. The latter value denotes the maximum CR confinement radius in our study. 

For our calculations in the following, we will assume that at the time of escape from the SNRs, CRs are distributed uniformly at the shock surface. This assumption is similar to that of the spherical surface source discussed in sections 2 and 3.

\subsection{CR proton spectrum from an SNR}
For CR protons, the source term components in the energy dependent escape model can be written as,
\begin{align}
& q(r_0)=\delta(r_0-R_{esc})\nonumber\\
& q(E)=\frac{q_p(E)}{A_{esc}}\nonumber\\
& q(t^\prime)=\delta(t^\prime-T_{esc})
\end{align}
where $q_p(E)$ is the point source spectrum given in section 2.1, $T_{esc}$ and $R_{esc}$ are given by Eq. (34) and Eq. (35) respectively and $A_{esc}=4\pi R_{esc}^2$. Now, the proton spectrum in this case is obtained using Eq. (13) by substituting the above source parameters as,
\begin{eqnarray}
N_p(E,t)=\frac{q_p(E)\,R_{esc}}{r_sA_{esc}\sqrt{\pi D(t-T_{esc})}}\mathrm{exp}\left[-\frac{\left(R_{esc}^2+r_s^2\right)}{4D(t-T_{esc})}\right]\nonumber\\
\times\;\mathrm{sinh}\left(\frac{r_sR_{esc}}{2D(t-T_{esc})}\right)
\end{eqnarray}

In Fig. 7, we show the proton spectra calculated using Eq. (37) for a source at $r_s=0.2$ kpc at different times $t=(10^3, 10^4, 10^5, 10^6)$ yr. The dashed lines correspond to $\alpha=1.1$ and the solid lines to $\alpha=2.0$. The sharp breaks in the spectra are due to the effect of the energy dependent escape time of the particles. Particles with energies below the breaks have not yet been escaped from the SNR or even if they do, they have not yet reached the Earth at the given time $t$. The effect of choosing different values of $\alpha$ is clearly visible. For $\alpha=1.1$, the spectra at all times except for $t\gg 10^5$ yr peak at relatively higher energies compared to those for $\alpha=2$. This is because particles at all energies except for those which remain till the end of the evolution are confined for relatively longer period in the case of $\alpha=1.1$ (see Fig. 6 top panel). Looking into the individual spectrum, we can also see that at high energies it follows a power-law spectrum as $N_p(E)\propto E^{-\left(\Gamma+\frac{3}{2}\delta\right)}$. This can be understood from Eq. (37) which shows that for particles with large  diffusion radius $r_{diff}\propto \sqrt{D(t-T_{esc}})$, the solution reduces to that of the point source approximation at high energies (Eq. 8). An additional effect of large $r_{diff}$ is that the spectra at the highest energies for the two different $\alpha$'s are very similar. These high energy particles are those which escaped the remnant long ago and their $r_{diff}\gg (R_{esc}, r_s)$ so that they have already passed the Earth at the given time. For these particles, the expected spectrum is almost independent of  the chosen values of $\alpha$, $T_{esc}$ and $R_{esc}$. This is more clearly visible in the results obtained for $t=10^6$ yr where the two spectra are almost identical to each other over the energy range considered here.

Although taking different values of $\alpha$ result into different types of spectrum especially at the lower energies at a given time, hereafter we will adopt $\alpha=2.0$ for our study. The effects of choosing other values of $\alpha$ on our results will be discussed later in section 6.

\begin{figure}
\centering
\includegraphics*[width=0.3\textwidth,angle=270,clip]{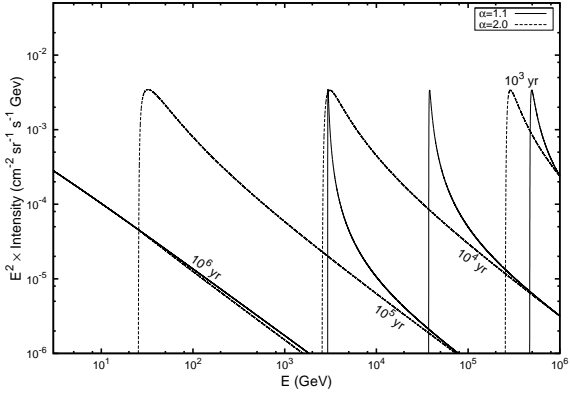}
\caption{\label {fig1} CR proton spectra at different times $t=(10^3, 10^4, 10^5, 10^6)$ yr under the energy dependent escape model for $\alpha=1.1$ (solid line) and $2.0$ (dashed line). We assume $r_s=0.2$ kpc, $\Gamma=2.0$, $D_0=2.9\times 10^{28}$ cm$^2$ s$^{-1}$, $\delta=0.6$, $E_0=3$ GeV, $E_{max}=10^6$ GeV and $t_{sed}=500$ yr.}
\end{figure}

\subsection{High energy electron spectrum from an SNR}
To proceed, we recall Eq. (26) which represents the electron spectrum $N_e(E)$ for the spherical surface source obtained under the energy independent escape model. In that equation, electrons of energy $E$ observe at time $t$ had an initial energy $E^\prime$ at the time of their escape given by,
\begin{equation}
E^\prime=\frac{E}{1-aE(t-t_0)}
\end{equation}
where $t_0$ denotes the escape time from the SNR. We can reverse the situation and calculate the energy of an electron after time $t$ for a given initial energy $E^\prime$ as,
\begin{equation}
E=\frac{E^\prime}{1+aE^\prime(t-t_0)}
\end{equation}
For an energy dependent escape, we can now substitute $t_0$ by $T_{esc}(E^\prime)$ and rewrite Eq. (39) as follows,
\begin{equation}
E=\frac{E^\prime}{1+aE^\prime\left[t-T_{esc}(E^\prime)\right]}
\end{equation}
If $q_e(E^\prime)$ represents the source spectrum of electrons with initial energy $E^\prime$ which escape the remnant at time $T_{esc}(E^\prime)$, their energy $E$ at time $t$ is given by Eq. (40) and their number density is obtained using Eq. (26) as given below,
\begin{eqnarray}
N_e(E,t)=\frac{q_e(E^\prime)\,R_{esc}^\prime}{r_s A_{esc}^\prime\sqrt{\pi C^\prime}}\left[1+aE^\prime(t-T_{esc}^\prime)\right]^2\qquad\qquad\nonumber\\
\qquad\qquad\qquad\times\;\mathrm{exp}\left[-\frac{\left(R_{esc}^{\prime 2}+r_s^2\right)}{4C^\prime}\right]\mathrm{sinh}\left(\frac{r_sR_{esc}^\prime}{2C^\prime}\right)
\end{eqnarray}
where $C^\prime$ is given by,
\begin{equation}
C^\prime=\frac{D(E^\prime)}{a(1-\delta)E^\prime}\left\lbrace 1-\left[1+aE^\prime(t-T_{esc}^\prime)\right]^{\delta-1}\right\rbrace
\end{equation}
and $R_{esc}^\prime\equiv R_{esc}(E^\prime)$, $T_{esc}^\prime\equiv T_{esc}(E^\prime)$ and $A_{esc}^\prime\equiv A_{esc}(E^\prime)$.

\begin{figure}
\centering
\includegraphics*[width=0.3\textwidth,angle=270,clip]{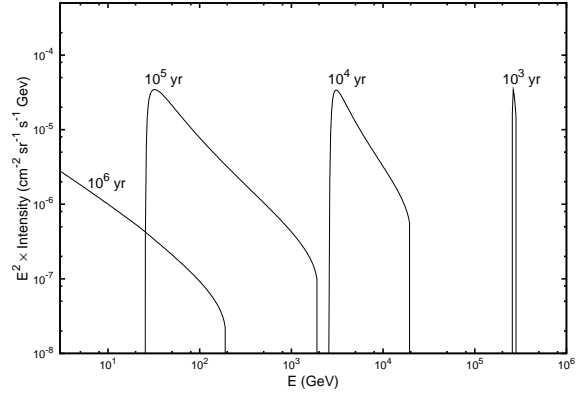}
\caption{\label {fig1} CR electron spectra at different times under energy dependent escape model for $\alpha=2.0$. Other model parameters are same as in Fig. 7.}
\end{figure}

Using Eq. (41), we calculate the electron spectra at different $t$'s for a source distance $r_s=0.2$ kpc. The results are shown in Fig. 8. On comparing with the results obtained for the protons shown in Fig. 7 $(\alpha=2.0)$, one can notice that the major difference is the presence of additional  breaks at higher energies which are due to the effect of radiative energy losses. The breaks at the lower energies which are due to the effect of $T_{esc}$ are seen at the same energies for both the type of particles. The electron spectrum between the breaks follow an exponent $\Gamma+\frac{3}{2}\delta$ similar to the proton spectrum and at very late times (say at $t=10^6$ yr), it also tends towards the point source solution. These results show that also in the case of energy dependent escape scenario, for very old sources ($t\gg 10^5$ yr) the spectrum at all energies can be well approximated by the simple point source solution.

\section{Application to the nearby SNRs}

\begin{table}
\centering
\caption{Parameters of known SNRs with distances $< 1$ kpc from the Earth. References: (1) Blair et al. 2005 (2) Tatematsu et al. 1990 (3) Leahy $\&$ Aschenbach 1996 (4) Leahy $\&$ Tian 2007 (5) Braun et al. 1989 (6) Caraveo et al. 2001 (7) Miceli et al. 2008 (8) Jian et al. 2005 (9) Strom 1994 (10) Plucinsky et al. 1996 (11) Caraveo et al. 1996 (12) Egger $\&$ Aschenbach 1995 (13) Katsuda et al. 2008 (where for the age we take the mean of $2700$ yr and $4300$ yr reported in the literature).}
\begin{tabular}{@{}lllrrlrlr@{}}
\hline
SNR			&	Distance (kpc)			&Age (yr)			&References\\         
\hline
Cygnus Loop 	&	0.540	&				$1.0\times 10^4$	&	1\\
HB21        		&	0.800	&				$1.9\times 10^4$	&	2, 3\\
HB9         		&	0.800	&				$6.6\times 10^3$	&	4\\
S147			&	0.800	&				$4.6\times 10^3$	&	5\\
Vela        		&	0.294	&				$1.12\times 10^4$	&	6, 7\\
G299.2-2.9		&	0.500	&				$5.0\times 10^3$	&	8\\
SN185			&	0.950	&				$1.8\times 10^3$	&	9\\
Monogem     	&	0.300	&				$8.6\times 10^4$	&	10\\
Geminga		&	0.157	&				$3.4\times 10^5$	&	11\\
Loop1			&	0.170	&				$2.0\times 10^5$	&	12\\
G114.3+0.3  		&	0.700	&				$4.1\times 10^4$	&	8\\
Vela Junior		&	0.750	&				$3.5\times 10^3$	&	13\\
\hline
\end{tabular}
\end{table}

In this section, we shall apply our study to the nearby known SNRs listed in Table 1 with distances $< 1$ kpc from the Earth. It should be mentioned that some of the age and the distance parameters listed in Table 1 carry large uncertainties. For instance, the distance to the Geminga was measured to be $157$ pc using Hubble Space Telescope (HST) observations (Caraveo et al. 1996) but recently, again using HST measurements, Faherty et al. 2007 reported the distance of Geminga to be $250$ pc. For the Cygnus Loop, Minkowski 1958 reported a distance of $770$ pc whereas measurements based on HST observations claimed a distance of $440$ pc (Blair et al. 1999). Recent measurements further claimed the distance to be $540$ pc (Blair et al. 2005). For HB21, Tatematsu et al. 1990 measured a distance of $800$ pc and Leahy $\&$ Aschenbach 1996 estimated an age of $1.9\times 10^4$ yr while later, Byun et al. 2006 suggested a distance of $1.7$ kpc and Lazendic $\&$ Slane 2006 estimated an age of $5.6\times 10^3$ yr. Leahy $\&$ Aschenbach 1995 estimated the distance and age of HB9 as $1$ kpc and $7.7\times 10^3$ yr respectively, and Leahy $\&$ Tian 2007 suggested a distance of $800$ pc with sedov age of $6.6\times 10^3$ yr  and age of $(4-7)\times 10^3$ yr based on evaporation cloud model. The lack of precise informations on these parameters can affect our results because of the strong dependence of the CR spectrum on these parameters.

For our study, we will assume that the proton source index $\Gamma=2.13$ so that for $\delta=0.6$, we get $\Gamma +\delta =2.73$ the observed proton spectral index at the Earth (Haino et al. 2004). It should be noted that the value of the source index can depend on the choice of the propagation model and different propagation models may take different values. For instance, models based on diffusive reacceleration in the Galaxy favors a diffusion index of $\delta\sim 0.3$ which corresponds to a source index of $\Gamma\sim 2.4$ (Trotta et al. 2011). This is steeper than the value adopted in our present work which is based on a purely diffusive model of CR propagation. For the CR electrons, to get the source index, we first determine the background spectrum. This is done by fitting the observed data between $(10-200)$ GeV provided by the FERMI and the PAMELA experiments (Ackermann et al. 2010b, Adriani et al. 2011). We assume that this is the energy region where the contamination due to the local sources as well as the effect of the solar modulation are minimum. From the fit, the background spectral index is found to be  $3.10\pm 0.01$. Under diffusive propagation model, CR electrons produced by a uniform and stationary source distribution, and subject to radiative losses during their propagation in the Galaxy results into an equilibrium spectrum given by $E^{-(\Gamma+1-\beta)}$ where $\beta=(1-\delta)/2$ (see e.g. Thoudam $\&$ H\"orandel 2011). Using the value of the background index obtained from the fit, we get the electron source index as $\Gamma=2.3$. This is the value we will adopt for the rest of our calculations for the electrons. Furthermore, in the following we will assume that $10\%$ of the supernova explosion energy of $10^{51}$ ergs converts into CR protons and $0.1\%$ into the electrons. All these parameters are assumed to be the same for all the SNRs.

\subsection{CR protons}
\begin{figure}
\centering
\includegraphics*[width=0.3\textwidth,angle=270,clip]{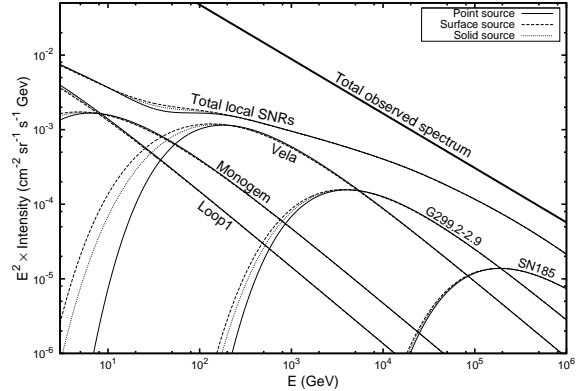}
\caption{\label {fig1} CR proton spectra from nearby SNRs listed in Table 1 for the three different source models: point source (solid line), surface source (dashed line) and solid source (dotted line). The contributions from the individual SNRs are labelled by their names and their total contributions as ``Total local SNRs". The thick solid line represents the fitted total observed spectrum, $1.37\times (E/\mathrm{GeV})^{-2.73}$ cm$^{-2}$ s$^{-1}$ sr$^{-1}$ GeV$^{-1}$ taken from Haino et al. 2004. Our calculation assumes an energy independent burst-like injection of particles at $t_0=0$ and that each SNR produces $10^{50}$ ergs of CR protons. Other model parameters: $\Gamma=2.13$, $D_0=2.9\times 10^{28}$ cm $^2$ s$^{-1}$, $\delta=0.6$ and $E_0=3$ GeV.}
\end{figure}
\begin{figure}
\centering
\includegraphics*[width=0.3\textwidth,angle=270,clip]{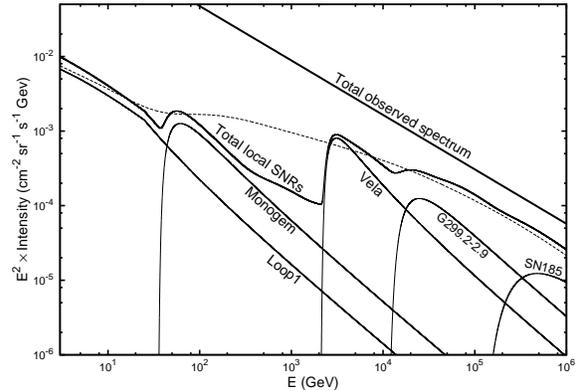}
\caption{\label {fig2} Same as in Fig. 9 but for an energy dependent escape model. We assume $\alpha=2.0$, $E_{max}=10^6$ GeV and $t_{sed}=500$ yr. The dashed line represents the ``Total local SNRs" we obtained in the case of the point source approximation shown in Fig. 9.}
\end{figure}
First, we compare the results of the point source approximation with those of the spherical solid and the surface source models which are all based on the energy independent escape model. These are shown in Fig. 9 where different lines represent different source model: thin solid (point), dashed (surface) and dotted (solid). The thick solid line represents the fitted observed proton spectrum given in Haino et al. 2004. The calculation assumes a burst-like injection of CRs at time $t_0=0$. The contributions from the dominant SNRs in different energy intervals are indicated by their names. Although some of the individual SNRs like the Vela and the G299.2-2.9 show different spectra at low energies under the different source models, the differences are not significant in the total combined spectrum from the nearby SNRs. It is because at low energies below $\sim 60$ GeV, the dominant contributions are from the Monogem and the Loop1 whose spectra does not show any differences between the models because of their old ages. Although different SNRs dominate at different energy intervals, as a whole their total spectrum looks smooth except for a slight dent somewhere between $\sim (10-100)$ GeV.

We have also checked the results for other values of $t_0\le10^5$ yr. Except for the signatures of the absence of young sources as $t_0$ takes larger values, we have found that their total spectra does not show any significant differences between the different source models at all values of $t_0$.

When we apply the energy dependent escape model, we find that the results are significantly different from those of the energy independent models. This is shown in Fig. 10. The dashed line represents the total spectrum we obtain for the point source approximation shown in Fig.9. We can see that the total spectrum in the energy dependent case  show irregular structures which are due to the low energy spectral breaks of the individual SNRs. However, such features can possibly remain embedded in the dominant CR background and may not be distinctly visible in the observed spectrum.

\subsection{Electrons}
The electron spectra for the point, solid and the surface source models are shown in Fig. 11. In the figure the data are  from the FERMI (Ackermann et al. 2010b), PAMELA (Adriani et al. 2011) and the HESS (Aharonian et al. 2008c, 2009) experiments. As in the case of protons, the total electron spectra also does not show any differences between the different models. However, unlike in the case of the protons, the total  electron spectra show some irregular features near the highest energies which are due to the effects of sharp cut-offs in the individual spectra due to radiative energy losses. For instance, the strong peak at $E\sim 10^5$ yr is due to the effect of SN185. From the figure, we can notice that at energies greater than few TeVs our results which are based on a pure power-law source spectrum significantly over predicts the data. Taking larger values of $t_0$ can suppress the contributions of Vela, G299.2-2.9 and SN185 which are the dominant contributors at high energies. For $t_0=2\times 10^4$ yr, their contributions will be completely removed. This points towards the importance of source modeling in order to understand the contribution of local sources in the high energy electron spectrum. One common way to handle this problem is to assume an exponential cut-off $\mathrm{exp}(-E/E_c)$ in the source spectrum at a few TeVs (see e.g. Delahaye et al. 2010). Another possibility is that the high energy electrons might have suffered significant energy losses within the SNR itself before they are released into the ISM (Thoudam $\&$ H\"orandel 2011). Therefore, electrons at higher energies might be released with a spectrum steeper than the lower energy ones. For the present study, we adopt the much simpler exponential cut-off and in Fig. 12, we show the results obtained for $E_c=2$ TeV. We can see that the shape of the total local spectrum in the TeV region is now determined mostly by the exponential cut-off and the irregular structures present near the highest energies in Fig. 11 no longer exist. In Fig. 12, the thick dashed line represents the background spectrum (which we obtain as mentioned before) with an index $3.1$ and an exponential cut-off at $2$ TeV. The thick solid line represents the total background plus the nearby SNRs contribution obtained in the point source approximation. Detailed calculations of the background spectrum taking into account the various source models discussed here will be presented elsewhere.

\begin{figure}
\centering
\includegraphics*[width=0.3\textwidth,angle=270,clip]{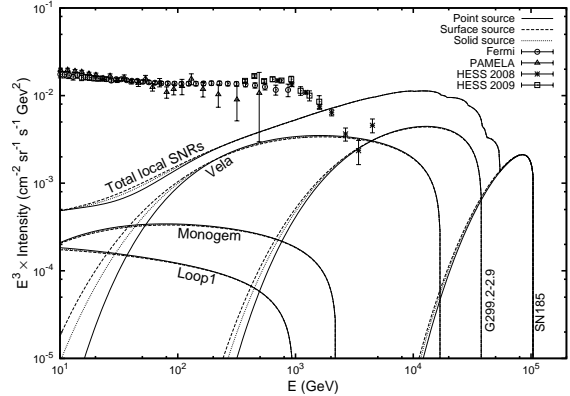}
\caption{\label {fig1} Electron spectrum from the nearby SNRs listed in Table 1 for the three different source models: point source (solid line), surface source (dashed line) and solid source (dotted line). We assume a burst-like injection of particles  and a pure power-law source spectrum of index $\Gamma=2.3$ with each SNR producing $10^{48}$ ergs of CR electrons. All other model parameters remain the same as in Fig. 9. The data are taken from the FERMI, PAMELA and HESS experiments.}
\end{figure}
\begin{figure}
\centering
\includegraphics*[width=0.3\textwidth,angle=270,clip]{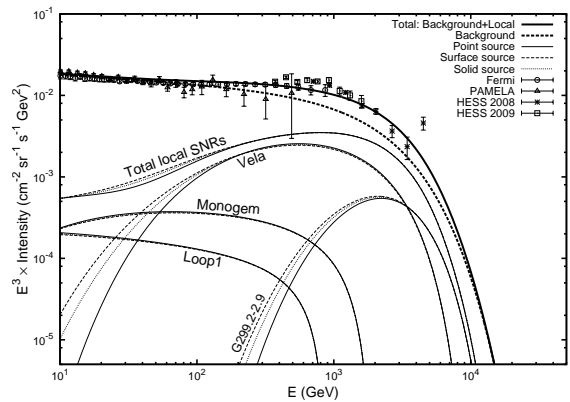}
\caption{\label {fig1} Same as in Fig. 11 but for a source spectrum with an exponential cut-off at $E_c=2$ TeV. The thick dashed line represents the background spectrum (see text for details) and the thick solid line represents the total background plus nearby SNRs in the point source approximation. The data are the same as in Fig. 11.}
\end{figure}
\begin{figure}
\centering
\includegraphics*[width=0.3\textwidth,angle=270,clip]{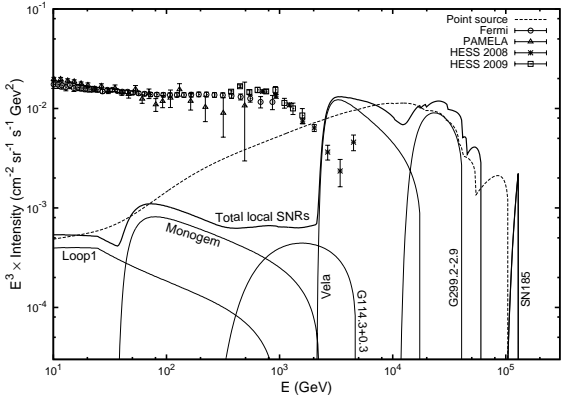}
\caption{\label {fig3} Electron spectrum from nearby SNRs for the energy dependent injection model and a pure power-law source spectrum. The dashed line represents the ``Total local SNRs" obtained under the point source approximation shown in Fig. 11. We assume $\alpha=2.0$, $E_{max}=10^6$ GeV and $t_{sed}=500$ yr. The data and all other model parameters remain the same as in Fig. 11.}
\end{figure}
\begin{figure}
\centering
\includegraphics*[width=0.3\textwidth,angle=270,clip]{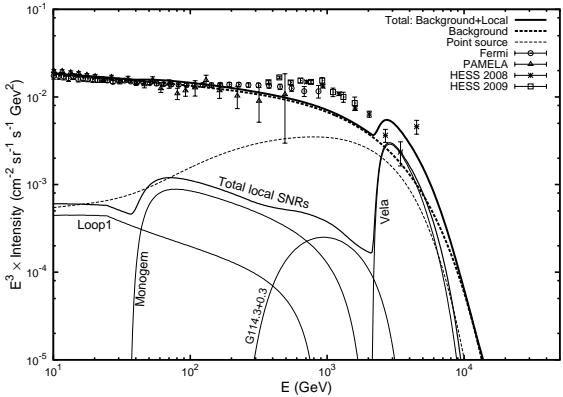}
\caption{\label {fig4} Same as in Fig. 13 but for a source spectrum with an exponential cut-off at $E_c=2$ TeV. The thick dashed line represents the background spectrum and the thin dashed line represents the ``Total local SNRs" for the point source approximation shown in Fig. 12. The data are the same as given in Fig. 11.}
\end{figure}

For the energy dependent escape model, the results are shown in Fig. 13 for a pure power-law source spectrum. In the figure, we also show for comparison the total local spectra obtained in the case of the point source approximation (dashed line in Fig.11). The total spectrum show several irregular features and spikes. These features are stronger than the ones present in the proton spectrum which is due to the presence of additional breaks in the individual electron spectra at high energies. The position of these spikes not only depends on the age and distance of the individual SNRs but also on the assume energy dependent escape model (i.e., on the parameters $\alpha$, $t_{sed}$ and $E_{max}$). 

In Fig. 14, we show the electron spectra obtained using the same model as in Fig. 13 but with an exponential cut-off in the source spectrum at $E_c=2$ TeV. In the figure, the thin and the thick dashed lines represent the local SNRs contribution in the case of the point source approximation and the background spectrum respectively as shown in Fig. 12. The thick solid line denotes the total background plus local spectrum for this case. We can notice that even after imposing the cut-off some prominent features still remain at few TeVs in the overall total spectrum unlike in the point source approximation where the cut-off almost smoothens the total spectrum.

\section{Overall results and discussions}
We have shown that the commonly adopted point source approximation does not always remain a good approximation as far as nearby CR sources are concerned. For a typical source distance of $r_s\sim (100-300)$ pc, we have shown that at low energies, the point source results for young sources (typically $t\lesssim 10^5$ yr) differ significantly from those calculated using a finite source size. At high energies the point source still remains a good approximation. Under the energy \textit{independent} particle escape model, we found that the effects of the finite source size are similar in both the types of particle injection model considered in our study: the burst-like and the continuous injection. For very old nearby sources ($t\gtrsim 10^5$ yr), we have found that the results are independent of both the source size and the particle injection model and hence, the burst-like point source model represents a good approximation at all energies. We have also shown in Fig. 3 that for a given value of the CR diffusion coefficient, there is a certain parameter space in $(r_s,t)$ under which the point source approximation remains valid for CRs of our interest, i.e., with energies $E>(3-10)$ GeV. When applied to the nearby known SNRs within $1$ kpc, interestingly we have found that their total spectrum almost remain the same in the three different source models although some of the individual SNRs like Vela show differences between the models (Figs. 9 $\&$ 11). We found that it is because at low energies where the point source approximation is most likely to break down, the local spectrum is dominated by the Monogem and the Loop1. These SNRs are quite old with Monogem age $\sim 8.6\times 10^4$ yr and Loop1 $\sim 2.0\times 10^5$ yr due to which their CR spectra at the Earth are independent of their sizes and are well represented by the point source solutions.

We have also studied an energy \textit{dependent} escape scenario where CRs of different energies are assumed to escape at different times during the SNR evolution. We assumed that the escape time follows, $t_{esc}\propto E^{-1/\alpha}$ with $\alpha$ chosen to be equal to $2.0$. Under this model, we assumed that the highest energy particles escape the remnant at the start of the Sedov phase followed by the lower energy ones at later times. For $E_{max}=1$ PeV, $t_{sed}=500$ yr and the maximum CR confinement time of  $10^5$ yr adopted for our study, we found $t_{esc}=(500-10^5)$ yr and the escape radius $R_{esc}=(5-100)$ pc for energies  $E=(1\mathrm{PeV}-25\mathrm{GeV})$. For young sources, the spectrum obtained under this model show breaks at lower energies which are due to the longer confinement times at those energies. At high energies, the results are very similar to those of the point source approximation. This is not just because of the small values of $R_{esc}$ at high energies but also due to their large values of $D(E)$ at these energies. In fact, we have shown in section 2 that even for a large escape radius of $100$ pc, the point source still represents a good approximation at high energies (see e.g., Fig. 1 surface source). Therefore, it should be understood that it is not the small $R_{esc}$, but actually the large $D(E)$ which is responsible for the point source validity at high energies under the energy dependent escape model. When applied to the nearby known  SNRs, we have found that the results obtained under this model are significantly different from those obtained under the point source approximation. The total local spectrum show more irregular structures as compared to the point source results. Also, we have noticed that there is a big dip between around $(10^2-3\times 10^3)$ GeV which is mainly due to the low energy cut-off in the Vela spectrum (Figs. 10 $\&$ 13). These results seem to  suggest that if SNRs are the main sources of CRs in our Galaxy, then the widely adopted point source approximation with an energy independent escape scenario appears flawed for CR studies from the nearby SNRs.

\begin{figure}
\centering
\includegraphics*[width=0.3\textwidth,angle=270,clip]{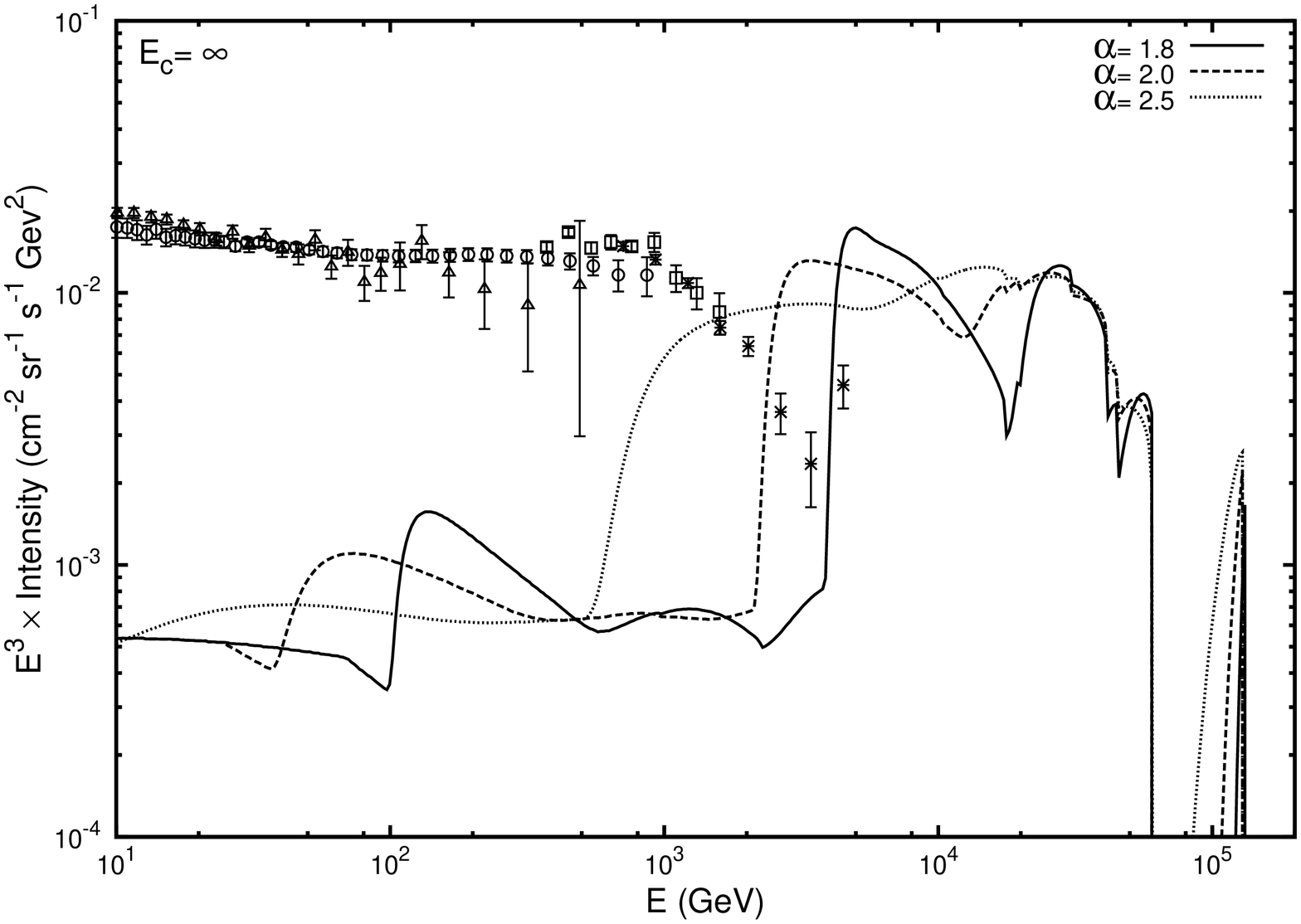}\\
\includegraphics*[width=0.3\textwidth,angle=270,clip]{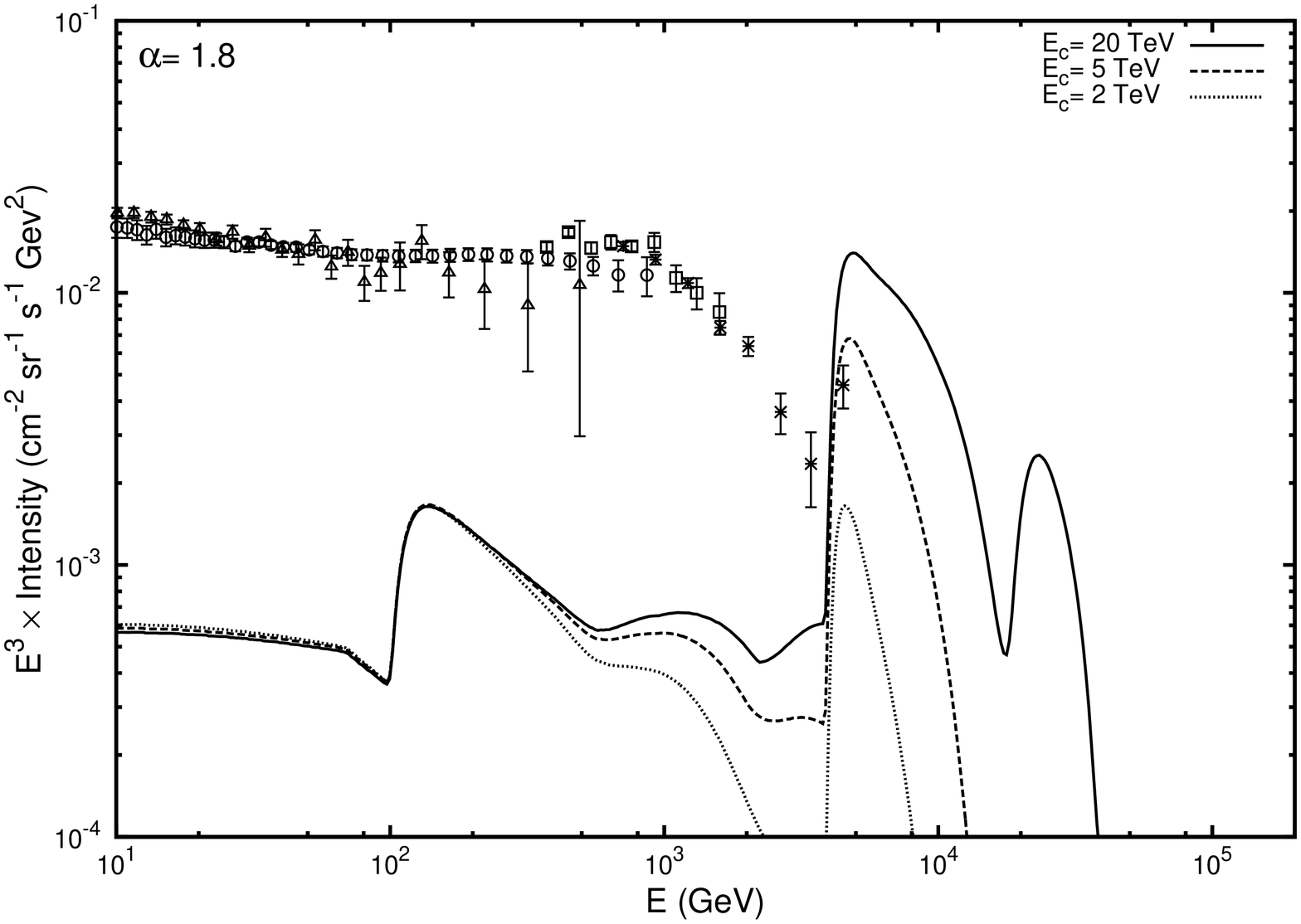}
\caption{\label {fig4} Top: Total electron spectrum from nearby SNRs assuming a pure power-law source spectrum $(E_c=\infty)$ for different values of $\alpha$: $1.8$ (solid line), $2.0$ (dashed line), $2.5$ (dotted line). Bottom: For $\alpha=1.8$ with exponential cut-off in the source spectrum at $E_c= 10$ TeV (solid line), $2$ TeV (dashed line) and $1$ TeV (dotted line). The data and other model parameters remain the same as in Fig. 11.}
\end{figure}

For the protons, the irregular spectral features that we have found in the energy dependent escape model may be suppressed by the dominant background produced by distant sources and hence, may not show up distinctly in the total observed spectrum. But, for the electrons they can possibly show up to detectable levels especially at TeV energies where the background level is expected to be significantly less. Recently, Kawanaka et al. 2011 proposed that such spectral features can be used to estimate the CR confinement time inside SNRs. Their study assumed a single nearby source having characteristics similar to that of the Vela remnant. It should be noted that the position and the strength of such features strongly depend on the CR escape model especially on the $\alpha$ parameter and also on $E_c$ (if there is an exponential cut-off in the source spectrum). For instance, assuming $E_c>2$ TeV would produce stronger features and vice versa compared to our results shown in Fig. 14. Similarly, assuming $\alpha<2.0$ would produce stronger peaks at comparatively higher energies as low energy CRs would be confined for relatively longer times as indicated by Figs. 6 $\&$ 7, and taking $\alpha>2.0$ would smoothen the peaks as low energy CRs would also start escaping at early times. For $\alpha\gg 2.0$, the energy dependent results will tend towards the point source results obtained for $t_0=t_{sed}$. These can be understood from Fig. 15 where in the top panel we have shown the  electron spectrum for the case of pure power-law source spectrum (which corresponds to $E_c=\infty$) for different values of $\alpha$: $1.8$ (solid line), $2.0$ (dashed line), $2.5$ (dotted line). We can clearly see the left shifting of the peak between $(1-10)$ TeV as $\alpha$ increases from $1.8$ to $2.5$. In the bottom panel, we have shown the spectra calculated for the case of $\alpha=1.8$ for source spectra with exponential cut-offs at $E_c=10$ TeV (solid line), $2$ TeV (dashed line) and $1$ TeV (dotted line). Here again, we can notice that the peak at around $5$ TeV grows stronger as $E_c$ takes larger values.
 
If we look into the HESS electron data, there is an indication of an abrupt rise at the highest measured energy. If future better sensitive experiments like the CTA and the CALET provide good quality data at these energies, that would indeed provide useful informations to understand CR escapes from some of our nearby SNRs. However, the large uncertainties involved in the age and the distance estimates of some of these SNRs may be an issue because of the strong dependence of the CR spectrum on these parameters. Regarding this, measurements of electron anisotropy both amplitude as well as its direction at these energies might also be important in order to identify the dominant source.

Recently, Di Bernardo et al. 2011 studied the contributions of the nearby pulsars and the SNRs to the high energy electron spectrum. One of their conclusions is that a strong contribution from the nearby SNRs is not supported by the recent upper limits on the electron anisotropies provided by the FERMI Large Area Telescope observations (Ackermann et al. 2010a). But, it should be noted that their calculations assumed the sources to be burst-like point sources emitting CR particles independent of energy. In Fig. 14, we show that between $\sim(100 \mathrm{GeV}- 2 \mathrm{TeV})$, the contribution from the nearby SNRs is significantly larger in the point source approximation than in the energy dependent model. 
We believe that a more realistic treatment of particle escape model from the SNRs may change their conclusion. Other class of sources which might also produce significant contributions to the high energy leptonic (electron plus positron) spectrum are pulsars and dark matter. Models based on these sources are motivated mostly by the  detection of the rise in the positron fraction above $\sim 10$ GeV by the PAMELA experiment (Adriani et al. 2009). If we assume that positrons are produced only during the interaction of the primary CRs with the interstellar gas, the positron fraction is expected to decrease with energy which is in contrast to the observations. A possible solution to this problem among others may be the presence of one or more nearby positron sources like pulsars or dark matter (see e.g. Grasso et al. 2009 and references therein). Future measurements of electron anisotropies with better sensitivities and also the absolute positron spectrum at high energies can provide better understanding of the nature and the type of the dominant source(s).

Moreover, a good understanding of the background contribution would also be crucial. In an earlier paper, we had presented calculations of the averaged background based on a simple energy \textit{independent} model of CR confinement within the SNRs (Thoudam $\&$ H\"orandel 2011). In  future, we will present background estimates for both the protons and the electrons taking into account the energy \textit{dependent} confinement/escape of particles. The calculation will include the various energy loss and the interaction processes taking place during the time particles are confined within the sources. In addition, we will also present the possible effects on other observed CR properties like the Galactic diffuse $\gamma$-ray emission, s/p ratios and the anisotropies.

\section*{Acknowledgments}
The authors would like to thank the anonymous referee for his/her constructive comments.
\\
\\
\textbf{REFERENCES}\\
\\
Abdo A. A. et al. 2009, ApJ, 706, L1\\
Abdo, A. A. et al. 2010a, ApJ, 712, 459\\
Abdo, A. A. et al. 2010b, ApJ, 718, 348\\
Abdo, A. A. et al. 2010c, Sci, 327, 1103\\
Ackermann, M. et al. 2010a, Phys. Rev. D 82, 092003\\
Ackermann, M. et al. 2010b, Phys. Rev. D 82, 092004\\
Adriani et al. 2009, Nature, 458, 607\\
Adriani et al. 2011, Phys. Rev. Lett. 106, 201101\\
Albert, J., et al. 2007, A$\&$A 474, 937\\
Aharonian,F. A., et al. 2006, ApJ, 636, 777\\
Aharonian, F. A., et al. 2008a, A$\&$A, 477, 353\\
Aharonian, F. A., et al. 2008b, A$\&$A, 481, 401\\
Aharonian, F. A., et al. 2008c, Phys. Rev. Lett., 101, 261104\\
Aharonian, F. A., et al. 2008d, A$\&$A, 490, 685\\
Aharonian, F. A., et al. 2009, A$\&$A, 508, 561\\
Aharonian, F.A., $\&$ Atoyan, A. 1996, A$\&$A, 309, 917\\
Atoyan, A. M., Aharonian, F. A., $\&$ V\"olk, H. J. 1995, Phys. Rev. D, 52, 3265\\
Bamba, A., Yamazaki, R., Yoshida, T., Terazawa, T., $\&$ Koyama, K. 2006, Adv. Space Res., 37, 1439\\
Beck, R. 2001, Space Science Reviews, 99, 243\\
Bell, A. R. 1978, MNRAS 182, 147\\
Bell, A. R. 2004, MNRAS, 353, 550\\
Berezhko, E. G., $\&$ V\"olk H. J. 2000, ApJ, 540, 923\\
Blair, W. P., Sankrit, R., Raymond, J. C., $\&$ Long, K. S. 1999, AJ, 118, 942\\
Blair, W. P., Sankrit, R., $\&$ Raymond, J. C. 2005, AJ, 129, 2268\\
Blandford, R., $\&$ Eichler, D. 1987, Physics Reports, 154, 1\\
Braun, R., Goss, W. M., $\&$ Lyne, A. G. 1989, ApJ, 340, 355\\
B\"usching, I., Kopp, A., Pohl, M., Schlickeiser, R., Perrot, C., $\&$ Grenier, I. 2005, ApJ, 619, 314\\
Byun, D.-Y., Koo, B.-C., Tatematsu, K., $\&$ Sunada, K. 2006, ApJ, 637, 283\\
Caprioli, D., Amato, E., $\&$ Blasi, P. 2010, Astropart. Phys.,33, 160\\
Caprioli, D., Blasi, P., $\&$ Amato, E. 2009, MNRAS, 396, 2065\\
Caraveo, P. A., Bignami, G. F., Mignami, R. P., $\&$ Taff, L. G. 1996, ApJ, 461, L91\\
Caraveo, P. A., De Luca, A., Mignani, R. P., $\&$ Bignami, G. F. 2001, ApJ, 561, 930\\
Casanova, S., et al. 2010, PASJ, 62, 1127\\
Delahaye, T., Lavalle, J., Lineros, R., Donato, F., Fornengo, N. 2010, A$\&$A, 524, 51\\
Di Bernardo, G., Evoli, C., Gaggero, D., Grasso, D., Maccione, L., Mazziotta, M. N., 2011, Astropart. Phys., 34, 528\\
Drury, L. O'C. 2011, 415, 1807\\
Egger, R. J., $\&$ Aschenbach, B. 1995, A$\&$A, 294, L25\\
Erlykin, A. D., Wolfendale, A. W. 2006, Astropart. Phys., 25, 183\\
Faherty, J., Walter, F. M., Anderson, J. 2007, Ap$\&$SS, 308, 225\\
Gabici, S., Aharonian, F. A., $\&$ Casanova, S. 2009, MNRAS, 396, 1629\\
Gaisser, T. K. 1990, Cosmic Rays and Particle Physics. Cambridge Univ. Press, Cambridge\\
Ginzburg, V. L., $\&$ Syrovatskii, S. I. 1964, The Origin of Cosmic Rays (Oxford: Pergamon)\\
Grasso, D., et al. 2009, Astropart. Phys., 32, 140\\
Gratton, L. 1972, ApSS, 16, 81\\
Green, D. A., 2009, BASI, 37, 45\\
Haino S. et al., 2004, Phys. Lett. B, 594, 35\\
Jian-Wen X., Xi-Zhen Z., $\&$ Jin-Lin H. 2005, Chin. J. Astron. Astrophys., 5, 165\\
Katsuda, S., Tsunemi, H., $\&$ Mori, K. 2008, ApJ, 678, L35\\
Kawanaka, N., Ioka, K., Ohira, Y., $\&$ Kashiyama, K. 2011, ApJ, 729, 93\\
Kobayashi, T., Komori Y., Yoshida K., $\&$ Nishimura J. 2004, ApJ, 601, 340\\
Lazendic, J. S., $\&$ Slane, P. O. 2006, ApJ, 647, 350\\
Leahy, D. A., $\&$ Aschenbach, B. 1995, A$\&$A, 293, 853\\
Leahy, D. A., $\&$ Aschenbach B., 1996, A$\&$A, 315, 260\\
Leahy, D. A., $\&$ Tian, W. W. 2007, A$\&$A, 461, 1013\\
Malkov, M. A., $\&$ Drury, L. O'C. 2001, Rep. Prog. Phys., 64, 429\\
Miceli, M., Bocchino, F., $\&$ Reale, F. 2008, ApJ, 676, 1064\\
Minkowski, R. 1958, Rev. Mod. Phys., 30, 1048\\
Ohira, Y., Murase, K., $\&$ Yamazaki, R. 2011, MNRAS, 410, 1577\\
Parizot, E., Marcowith, A., Ballet, J., $\&$ Gallant, Y. A. 2006, A$\&$A, 453, 387\\
Plucinsky, P. P., et al. 1996, ApJ, 463, 224\\
Porter, T. A., Moskalenko, I. V., Strong, A. W., Orlando, E., $\&$ Bouchet, L., 2008, ApJ, 682, 400\\
Ptuskin, V. S., $\&$ Zirakashvili, V. N. 2005, A$\&$A, 429, 755\\
Shen, C. S. 1970, ApJ, 162, L181\\
Shibata, T., Ishikawa, T., $\&$ S. Sekiguchi, S., 2011, ApJ, 727, 38\\
Stephens, S. A., $\&$ Streitmatter, R. E. 1998, ApJ, 505, 266\\
Strom, R. G. 1994, A$\&$A, 288, L1\\
Strong, A. W., $\&$ Moskalenko, I. V. 2001, 27th ICRC, Hamburg, Germany,1942\\
Taillet, R., Salati, P., Maurin, D., Vangioni-Flam, E., $\&$ Cass$\mathrm{\acute{e}}$, M. 2004, ApJ, 609, 173\\
Tatematsu, K., Fukui, Y., Landecker, T. L., $\&$ Roger, R. S. 1990, A$\&$A, 237, 189\\
Thoudam, S. 2007, MNRAS Letters, 380, L1\\
Thoudam, S. 2008, MNRAS, 388, 335\\
Thoudam, S., $\&$ H\"orandel, J. R. 2011, MNRAS, 414, 1432\\
Trotta, R., J\'ohannesson, G., Moskalenko, I. V., Porter, T.A., Ruiz de Austri, R., $\&$ Strong, A. W., 2011, ApJ, 729, 106\\
V\"olk, H. J., Berezhko, E. G., $\&$ Ksenofontov, L. T. 2005, A$\&$A, 433, 229\\
\end{document}